\documentclass[twocolumn,trackchanges]{aastex631}

\usepackage{amsmath}
\usepackage{amssymb}
\usepackage{hyperref}
\usepackage{natbib}
\usepackage{enumitem}
\usepackage[mathlines]{lineno}

\newcommand{\Teff}{T_{\rm eff}}
\newcommand{\Acc}{A_{\text{cc}}}
\newcommand{\AIa}{A_{\text{Ia}}}
\newcommand{\qcc}{q_{\text{cc}}}
\newcommand{\qIa}{q_{\text{Ia}}}

\defcitealias{Weinberg2022}{W22}
\defcitealias{Sit2024}{S24}
\defcitealias{Griffith2023}{G24}

\begin{document}

\title{On the Origin of Abundance Variations in the Milky Way's High-$\alpha$ Plateau}

\correspondingauthor{Tawny Sit}
\email{sit.6@osu.edu}

\author[0000-0001-8208-9755]{Tawny Sit}
\affiliation{Department of Astronomy and Center of Cosmology and AstroParticle Physics, The Ohio State University, Columbus, OH 43210, USA}

\author[0000-0001-7775-7261]{David H. Weinberg}
\affiliation{Department of Astronomy and Center of Cosmology and AstroParticle Physics, The Ohio State University, Columbus, OH 43210, USA}

\author[0000-0001-9345-9977]{Emily J. Griffith}
\altaffiliation{NSF Astronomy and Astrophysics Postdoctoral Fellow}
\affiliation{Center for Astrophysics and Space Astronomy, Department of Astrophysical and Planetary
Sciences, University of Colorado, 389 UCB, Boulder, CO 80309-0389, USA}

\begin{abstract}

Using multi-element abundances from the SDSS APOGEE survey, we investigate the origin of abundance variations in Milky Way (MW) disk stars on the ``high-$\alpha$ plateau," with $-0.5\leq\rm{[Mg/H]}\leq-0.1$ and $0.25\leq\rm{[Mg/Fe]}\leq0.35$. The elevated [$\alpha$/Fe] ratios of these stars imply low enrichment contributions from Type Ia supernovae (SNIa), but it is unclear whether their abundance patterns reflect pure core-collapse supernova (CCSN) enrichment. We find that plateau stars with higher [Fe/Mg] ratios also have higher [X/Mg] ratios for other iron-peak elements, suggesting that the [Fe/Mg] variations in the plateau population do reflect variations in the SNIa/CCSN ratio. To quantify this finding, we fit the observed abundance patterns with a two-process model, calibrated on the full MW disk, which represents each star's abundances as the sum of a prompt CCSN process with amplitude $\Acc$ and a delayed SNIa process with amplitude $\AIa$. This model is generally successful at explaining the observed trends of [X/Mg] with $\AIa/\Acc$, which are steeper for elements with a large SNIa contribution (e.g., Cr, Ni, Mn) and flatter for elements with low SNIa contribution (e.g., O, Si, Ca). Our analysis does not determine the value of [Mg/Fe] corresponding to pure CCSN enrichment, but it should be at least as high as the upper edge of the plateau at $\rm{[Mg/Fe]}\approx0.35$, and could be significantly higher. Compared to the two-process predictions, the observed trends of [X/Mg] with $\AIa/\Acc$ are steeper for (C+N) but shallower for Ce, providing intriguing but contradictory clues about AGB enrichment in the early disk.

\end{abstract}

\section{Introduction} \label{sec:intro}

Stars in the Milky Way (MW) disk follow two distinct sequences in the plane of [$\alpha$/Fe] vs. [Fe/H], with the ``high-$\alpha$" and ``low-$\alpha$" populations associated (though not perfectly: see \citealt{Hayden2017}) with the kinematic hot, thick disk and the cold, thin disk, respectively. Low-$\alpha$ stars are generally younger than high-$\alpha$ stars of the same metallicity, and the lower (around solar) [$\alpha$/Fe] is understood as a consequence of greater Fe enrichment from time-delayed Type Ia supernovae (SNIa) \citep[e.g.,][]{Matteucci1986,McWilliam1997,Fuhrmann1998,Bensby2003,Adibekyan2012}. While the ``knee" where the [$\alpha$/Fe] ratio begins to decrease with increasing metallicity is a signature of SNIa enrichment in the thick disk \citep[e.g.,][]{Feltzing2003}, a common assumption is that the [$\alpha$/Fe] ratios in the high-$\alpha$ populationat metallicities below the knee, where the [$\alpha$/Fe] trend is mostly flat, reflect the yield ratios for pure core collapse supernova (CCSN) enrichment, averaged over the stellar initial mass function (IMF) and other progenitor properties like rotation and binarity. Alternatively, the higher [$\alpha$/Fe] ratio could represent a mix of SNIa and CCSN contributions, but with a SNIa/CCSN ratio lower than that of the low-$\alpha$ population. In either case, it is more physically accurate to describe the ``high-$\alpha$" and ``low-$\alpha$" populations as ``low-Ia" and ``high-Ia," respectively, which is the nomenclature we adopt in the rest of this paper. Using data from the SDSS APOGEE survey \citep{Majewski2017}, \citet{deLis2016} estimate an intrinsic scatter of $\sim$0.04 dex in [O/Fe] at fixed [Fe/H] {\it within} the low-Ia population. Similarly, \citet{Vincenzo2021} estimate intrinsic scatter of $\sim$0.03-0.04 dex in the low-Ia population for [O/Fe], [Mg/Fe], [Si/Fe], and [Ca/Fe].

In this paper, we investigate the abundance scatter of 17 elements within the low-Ia population. One key question is whether abundance scatter within the low-Ia population is driven primarily by variations in the SNIa/CCSN ratio or arises instead from other effects such as stochastic sampling of the supernova population. If there are significant variations of SNIa/CCSN within the low-Ia population, then elements that have a large SNIa contribution should vary together. For example, stars that are high in [Fe/Mg] should also be high in [Ni/Mg] and [Mn/Mg], where we have adopted Mg as our reference element because it is expected to arise entirely from CCSN with little metallicity dependence \citep[see, e.g.,][and references therein]{Nomoto2013,Andrews2017}. Conversely, elements whose production is dominated by CCSN should exhibit low scatter in [X/Mg] at fixed [Mg/H]. 

To implement this idea quantitatively, we adopt the two-process model \citep{Weinberg2019,Weinberg2022,Griffith2019,Griffith2022,Griffith2023,Sit2024}, which describes a star's multi-element abundances as the sum of enrichment from a prompt process associated with CCSN and a delayed process associated with SNIa. The two-process model has been successful at predicting the multi-element abundance pattern of stars in the MW disk to $\approx$0.02$-$0.05 dex accuracy, comparable to the observational uncertainties reported by large spectroscopic surveys. At each [Mg/H], the relative contribution of the two processes for a given element X is inferred from the gap in median [X/Mg] between the disk's low-Ia and high-Ia populations. If abundance scatter in the low-Ia population comes mainly from SNIa/CCSN variation, then the two-process model should predict the element-by-element trends \textit{within} this population, even though it is calibrated to the difference between low-Ia and high-Ia stars. However, the two-process model is limited by its namesake two processes, SNIa and CCSN, so deviations can be indicative of other nucleosynthetic pathways, such as enrichment from asymptotic giant branch (AGB) stars \citep{Griffith2022,Weinberg2022}. Therefore, we can also investigate the contribution of any additional processes to the abundance scatter in the low-Ia plateau.

We concentrate our analysis on low-Ia stars with $-0.5 \leq \rm{[Mg/H]} \leq -0.1$, a range where the median trend of [Mg/Fe] with [Mg/H] is well described by a flat ``plateau" at [Mg/Fe]$_{\rm pl} \approx 0.3$. Many observational studies show this plateau continuing down to $\rm{[Fe/H]} \approx -2.5$, with similar flat plateaus for [Si/Fe] and [Ca/Fe] \citep[see, e.g.,][]{Kobayashi2020}. It is tempting, therefore, to identify these plateau ratios with the ratios produced by CCSN alone. However, the star-to-star scatter and study-to-study variations become larger at $\rm{[Fe/H]} < -1$. Furthermore, \cite{Conroy2022} find that {\it in situ} halo stars show a trend of declining [Mg/Fe] with increasing [Fe/H] for $\rm{[Fe/H]} < -1.5$, rather than a flat plateau. They propose a model in which the true ratio from CCSN enrichment corresponds to [Mg/Fe]$_{\rm cc} \approx 0.6$, and the roughly flat trend of [Mg/Fe] in the thick disk reflects a balance between SNIa and CCSN enrichment during an epoch of accelerating star formation \citep[see also][]{Maoz2017,Chen2023}.

The two-process model recasts each star's abundances into two amplitude parameters corresponding to the level of prompt (CCSN) and delayed (SNIa) contribution, and the element-by-element residuals between the observed abundances and the model predictions. Both the ``process vectors" describing the relative element contributions and the inferred values of the star-by-star ``process amplitudes" depend on the value of [Mg/Fe] that is assumed to represent pure CCSN enrichment (see Section \ref{sec:2proc} below). We originally hoped that our analysis could distinguish a scenario in which the low-Ia plateau reflects CCSN only from a scenario in which SNIa enrichment is already substantial at $\rm{[Mg/Fe]} \approx 0.3$.  However, we show in Section \ref{subsec:qs_As} that the residual abundances from the two-process predictions are independent of the assumed [Mg/Fe]$_{\rm cc}$, even though the process amplitudes are not (see also \citealt{Griffith2023}). Unfortunately, this means that our results do not shed light on the true level of [Mg/Fe]$_{\rm cc}$, though they do provide strong evidence for variation of the SNIa/CCSN ratio within the low-Ia population.

This paper is organized as follows: Section \ref{sec:data} introduces the abundance data we use. In Section \ref{sec:2proc}, we discuss the two-process model in detail, showing that it predicts abundances through a simple linear interpolation of median \textit{linear} abundance ratios and that the predicted abundances are independent of the assumed [Mg/Fe]$_{\rm cc}$ as a result. Section \ref{sec:scatter} examines the sources of observed scatter around the median [X/Mg] values in the low-Ia plateau, where we find that variation in SNIa/CCSN contributes strongly to the observed intrinsic dispersion. In Section \ref{sec:conclusions} we summarize our results and discuss future avenues for studying the information encoded in the intrinsic dispersion patterns of abundances in the Galaxy.

\section{Data}\label{sec:data}

Systematic trends in abundance measurements with $\log(g)$ and $\Teff$ can cause changes in median abundance trends \citep{Griffith2021} or, more importantly for this study, add correlated scatter between elements \citep{Weinberg2022}. These trends are most likely an artifact of the spectral fitting process, as a star's $\log(g)$ changes during its post-main sequence lifetime, but the abundances of most elements should stay the same. \citet[hereafter \citetalias{Sit2024}]{Sit2024} present corrected abundances for 288,789 stars from APOGEE DR17 \citep{Abdurrouf2022}, where artificial trends in APOGEE abundances with stellar surface gravity $\log(g)$ were identified and removed. This $\log(g)$ calibration also removes trends with stellar effective temperature $\Teff$.

In this work, we use a calibration sample and a smaller plateau sample, both selected from the full \citetalias{Sit2024} catalog. Our calibration sample is the same as that of \citetalias{Sit2024}, whose selection criteria are: 
\begin{itemize}[noitemsep]
    \item $-0.75\leq\rm{[Mg/H]}\leq0.45$
    \item $\log(g) = 0-3.5$
    \item $\Teff = 3000-5500$ K
    \item $R = 3-15$ kpc
    \item $|Z| < 2$ kpc
    \item \texttt{EXTRATARG==0}
    \item SNR $\geq$ 100
\end{itemize}
The calibration sample consists of 149,904 stars from APOGEE DR17.

\begin{figure*}[!ht]
    \centering
    \includegraphics[width=\textwidth]{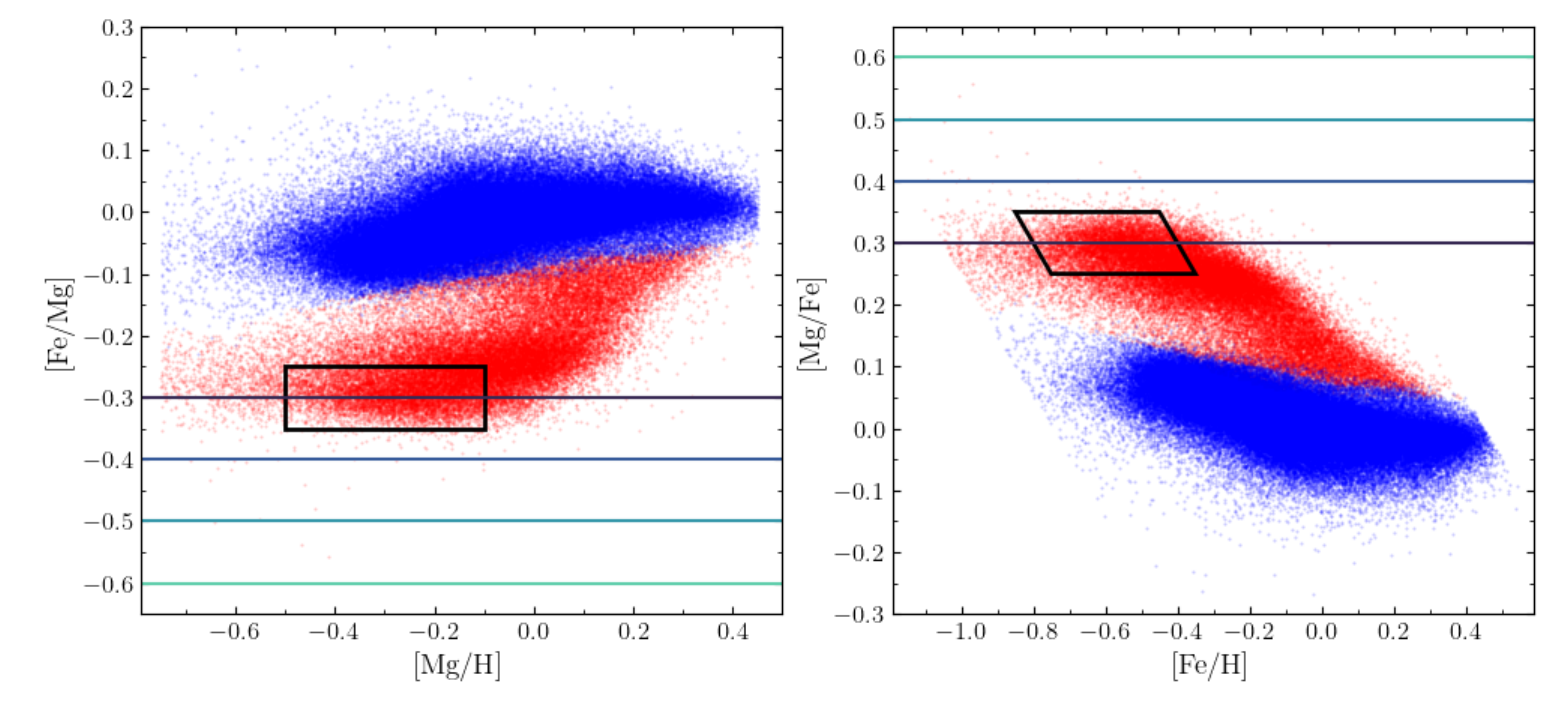}
    \caption{Low-Ia (red) and high-Ia (blue) APOGEE DR17 stars from the calibration sample of the \citetalias{Sit2024} catalog in[Fe/Mg]-[Mg/H] space (left) and [Mg/Fe]-[Fe/H] space (right). Plotted abundances are the \citetalias{Sit2024} $\log(g)$-calibrated abundances. The sample selection criteria (see Section \ref{sec:data}) for low-Ia plateau stars (``plateau sample") analyzed in this work are indicated by the black lines. The four different [Fe/Mg]$_{\rm cc}$ assumptions ([Fe/Mg]$_{\rm cc} = -0.3, -0.4, -0.5, -0.6$) tested in Section \ref{sec:2proc} are indicated by the horizontal lines.}
    \label{fig:sample}
\end{figure*}

We also generate a smaller plateau sample of stars that occupy a well-populated region on the low-Ia plateau and is a subset of the calibration sample. We apply the following cuts in [Fe/Mg] and [Mg/H] to the calibration sample to define the plateau sample:
\begin{itemize}[noitemsep]
    \item $-0.5 \leq \rm{[Mg/H]} \leq -0.1$
    \item $-0.35 \leq \rm{[Fe/Mg]} \leq -0.25$
\end{itemize}
These cuts select the region with the highest density of plateau stars around the [Mg/Fe]$_{\rm cc}$ = 0.3 value used in \citet{Weinberg2022} and \citetalias{Sit2024}. The plateau sample contains 10,262 stars.  

Figure \ref{fig:sample} shows the calibration sample, divided into low-Ia and high-Ia populations, and the additional [Mg/Fe] and [Fe/H] cuts that define the plateau sample. The low-Ia population is defined as
\begin{equation}
    \begin{cases}
    \text{[Mg/Fe]} > 0.12 - 0.13\text{[Fe/H]}, & \text{[Fe/H]} < 0\\
    \text{[Mg/Fe]} > 0.12, & \text{[Fe/H]} > 0.
\end{cases}
\end{equation}
For this division between the low-Ia and high-Ia populations \textit{only}, we use the raw ASPCAP values (i.e., \textit{prior} to the $\log(g)$ calibration). This maintains consistency with \citetalias{Sit2024} in the median [X/Mg] sequences of each population from which the ``process vectors" of the two-process model are derived (see Section \ref{sec:2proc}), so that we better isolate the effect of varying the assumed core-collapse [Fe/Mg] ratio in the two-process model in Section \ref{sec:2proc}. After accounting for flat offsets applied during calibration (see \citetalias{Sit2024} for details), applying the $\log(g)$ calibrations would change the high/low-Ia classification of only 1\% of the calibration sample, and only for stars near the division boundary (none of which are in the plateau sample).

We utilize 17 elements, also analyzed by \citetalias{Sit2024}, spanning a range of nucleosynthetic sources: O, Si, S, and Ca ($\alpha$ elements); C+N, Na, Al, and K (light odd-Z elements); Cr, Fe, Ni, V, Mn, Co, and Cu (iron-peak elements), and Ce ($s$-process element). The combined element C+N accounts for the surface abundances of C and N changing as a star evolves up the red giant branch while the total number of nuclei remains nearly equal to the birth abundance. \citetalias{Sit2024} incorporated abundance measurements from the BAWLAS catalog \citep{Hayes2022} into their analysis for O, C+N, S, Na, V, and Ce, but we use the ASPCAP measurements for this work to increase the sample size of available measurements, since BAWLAS used a higher SNR threshold of SNR $\geq$ 150. Cu is on the sharply falling edge of the iron peak (``Fe-cliff") and though its astrophysical origins are debated, literature suggests more significant contributions from massive stars \citep[e.g.,][]{Woosley1995,Romano2007,Kobayashi2020}, so it may have different nucleosynthesis patterns from the other Fe-peak elements. Therefore, despite being measured only by BAWLAS, we proceed with including Cu in this study, but note that only 2,277 out of 10,262 stars in our sample have a Cu measurement.

As part of our analysis in this paper, we use the reported measurement errors from ASPCAP (or BAWLAS, for Cu). The measurement uncertainties in DR17 are based on the process for DR16, where a function of $\Teff$, [M/H], and SNR is fit to the deviations in measured abundances from repeat observations (see Section 5.4 of \citealt{Jonsson2020} for more details). A similar procedure using repeat measurements is used for the empirical BAWLAS uncertainties (see Section 4.7.2 in \citealt{Hayes2022}) we use for Cu.

\section{The Two-Process Model}\label{sec:2proc}

The two-process model is a semi-empirical nucleosynthesis model that describes a star's multi-element abundances as a sum of two contributions, associated with CCSN and SNIa, with amplitudes $\Acc$ and $\AIa$. The processes are anchored in observed median [X/Mg] trends with metallicity, which have been observed to be nearly independent of location across the Galactic disk provided the low-Ia and high-Ia populations are separated \citep{Weinberg2019}. By interpreting these spatially invariant trends as driven by nucleosynthetic yields rather than spatially-varying quantities such as the star formation history, the relative contribution to different elements is taken to be universal for each process (though it may depend on metallicity). These relative contributions reflect the population-averaged yields of CCSN and SNIa, where the population-average includes averages over the stellar IMF, binary properties, rotation, and other parameters that affect yields. In the notation of \citet[hereafter \citetalias{Weinberg2022}]{Weinberg2022}, the abundances predicted by the two-process model are
\begin{equation}\label{eq:2proc_def}
    {\rm [X/H]_{2proc}} = \log_{10}\left(\Acc \qcc^X + \AIa \qIa^X\right),
\end{equation}
where the ``process vectors" $\qcc^X$ and $\qIa^X$ are derived from the ensemble of stars and the amplitudes $\Acc$ and $\AIa$ are fit to each star individually. 

The two-process model can successfully predict stellar abundances in the MW disk to 0.02-0.05 dex on average (e.g., \citealt{Weinberg2019}; \citealt{Griffith2019}; \citetalias{Weinberg2022}), and similar accuracy has been achieved with other two-parameter models conditioned on age and [Fe/H] \citep{Ness2019} or $\alpha$ and Fe abundances \citep{Ness2022,Ratcliffe2023}. However, a star's actual abundances may differ from the two-process prediction because additional processes contribute or because the CCSN and SNIa processes are not completely universal. The approach can be generalized to include additional processes (\citetalias{Weinberg2022}; \citealt[hereafter \citetalias{Griffith2023}]{Griffith2023}), though these become more difficult to disentangle as the number of processes increases. More generally, one can think of the two processes as representing prompt and time-delayed nucleosynthesis contributions, and the two-process model is reasonably successful at predicting abundances of elements whose delayed contribution is expected to arise from asymptotic giant branch (AGB) stars rather than SNIa, such as C, N, Ce, Y, and Ba (\citealt{Griffith2022}; \citetalias{Weinberg2022}).

\subsection{Process Vectors and Amplitudes} \label{subsec:qs_As}
Without loss of information, one can express a star's measured abundances in terms of the two-process amplitudes $\Acc$ and $\AIa$, which characterize the overall level of prompt and delayed enrichment, and residuals from the two-process prediction
\begin{equation}\label{eq:resid_abund}
    \Delta\rm{[X/H]_{\rm 2proc}} = \rm{[X/H]_{meas}} - \rm{[X/H]_{2proc}}.
\end{equation}

For present purposes, we would like to know how the predicted and residual abundances depend on the value [Fe/Mg]$_{\rm cc}$ that is assumed to represent the iron-to-magnesium ratio from CCSN enrichment with no SNIa contribution. We test assumptions of [Fe/Mg]$_{\rm cc}$ values spaced at 0.1 dex intervals between [Fe/Mg]$_{\rm cc} = -0.3$, the fiducial assumption from \citetalias{Weinberg2022} and \citetalias{Sit2024}, and [Fe/Mg]$_{\rm cc} = -0.6$, approximately the value estimated by \citet{Conroy2022}. We follow the definition and equations of \citetalias{Weinberg2022}, and to simplify expressions we introduce the notation
\begin{equation}
    \{\text{X/Y}\} = 10^{\text{[X/Y]}},
\end{equation}
i.e., \{X/Y\} is simply the linear abundance ratio of elements X and Y scaled by the solar ratio. \citetalias{Weinberg2022} derive the two-process vectors from the median abundance ratio of stars in the low-Ia and high-Ia disk populations, [X/Mg]$_{\text{low}}$ and [X/Mg]$_{\text{high}}$ respectively, in bins of [Mg/H].\footnote{\citetalias{Griffith2023} present an alternative method that derives $\qcc^X$ and $\qIa^X$ from global fits to the ensemble of stellar abundances; the two approaches give very similar results.} They assume that Mg comes entirely from CCSN and use \{Fe/Mg\} as an indicator of the amount of SNIa enrichment. 

\begin{figure*}[!th]
    \centering
    \includegraphics[width=\textwidth]{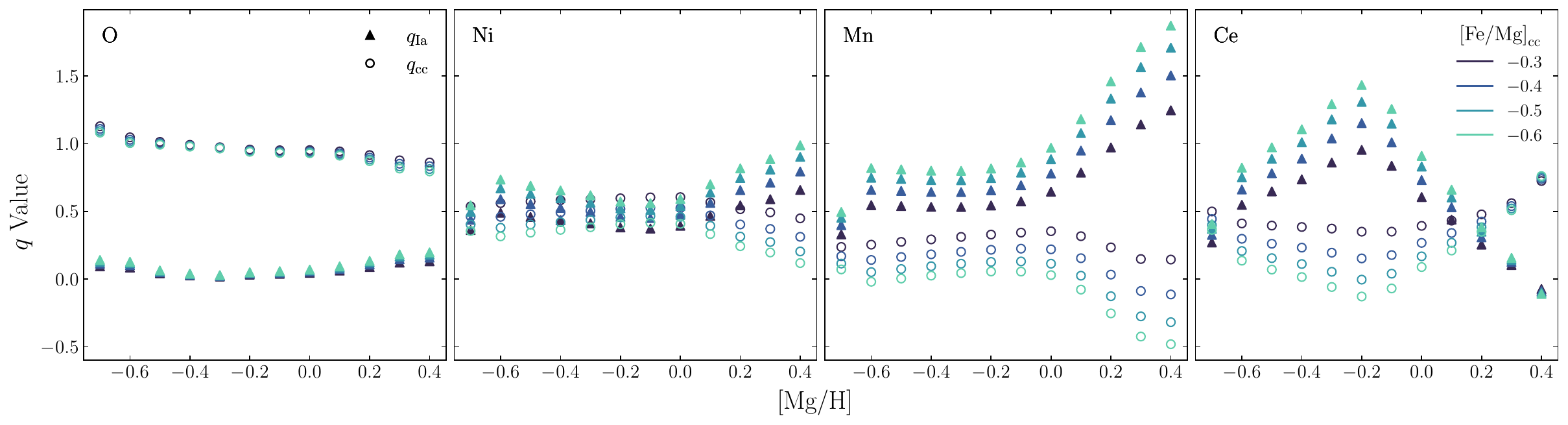}
    \caption{$\qIa^X$ (filled triangles) and $\qcc^X$ (open circles) vectors as a function of metallicity for X = O, Ni, Mn, and Ce, derived from the full \citetalias{Sit2024} calibration sample. Colors indicate different assumed plateau values, with increasingly light colors reflecting a decreasing [Fe/Mg]$_{\rm cc}$ value (or increasing [Mg/Fe]$_{\rm pl}$); the colors match the lines depicted in Figure \ref{fig:sample}.}
    \label{fig:q_arrays}
\end{figure*}

One can use equations 21-26 of \citetalias{Weinberg2022} to show that
\begin{multline}\label{eq:qcc}
    \qcc^X = \rm{\{X/Mg\}_{low}} - \rm{\left(\{X/Mg\}_{high} - \{X/Mg\}_{low}\right)}\\
    \times \left(\frac{\rm{\{Fe/Mg\}_{low}-\{Fe/Mg\}_{cc}}}{\rm{\{Fe/Mg\}_{high}-\{Fe/Mg\}_{low}}}\right)
\end{multline}
and
\begin{multline}\label{eq:qIa}
    \qIa^X = \rm{\left(\{X/Mg\}_{high} - \{X/Mg\}_{low}\right)}\\
    \times\left(\frac{\rm{1-\{Fe/Mg\}_{cc}}}{\rm{\{Fe/Mg\}_{high}-\{Fe/Mg\}_{low}}}\right).
\end{multline}
Expressed in these linear, solar-scaled ratios, the value of $\qIa^X$ follows from the gap in \{X/Mg\} between the high-Ia and low-Ia populations, divided by the gap in \{Fe/Mg\} between these populations relative to the total gap between the CCSN ratio and the solar ratio. If \{X/Mg\} is the same for low-Ia and high-Ia stars, then $\qIa^X=0$, and $\qcc^X$ is simply the solar-scaled \{X/Mg\}. Equations \ref{eq:qcc} and \ref{eq:qIa} are applied in each [Mg/H] bin. The assumed value of \{Fe/Mg\}$_{\text{cc}}$ can in principle change with [Mg/H], though \citetalias{Weinberg2022} adopt a single value based on the plateau in [Mg/Fe] observed for low-metallicity, low-Ia disk stars, while \citetalias{Griffith2023} include a possible tilt in this plateau.

Figure \ref{fig:q_arrays} plots $\qIa^X$ and $\qcc^X$ as a function of [Mg/H] with different [Fe/Mg]$_{\rm cc}$ assumptions for four illustrative elements. $\qIa^X$ and $\qcc^X$ are derived from the calibration sample described in Section \ref{sec:data}. Keeping in mind that the two-process model, as defined by \citetalias{Weinberg2022}, fixes $\qIa^{\rm Mg} = 0$ and $\qcc^{\rm Mg}=1$ irrespective of the assumed plateau value, we observe that changing the [Fe/Mg]$_{\rm cc}$ assumption does not significantly change $\qIa^{\rm O}$ and $\qcc^{\rm O}$ because O is an $\alpha$ element like Mg. The other elements illustrated here (Ni, Mn, and Ce), all have a significant delayed contribution from SNIa (Ni and Mn) or AGB stars (Ce). For these elements, $\qIa^{\rm X}$ increases with decreasing [Fe/Mg]$_{\rm cc}$, while $\qcc^{\rm X}$ decreases; however, the overall shape of the metallicity-dependent $q$-vector stays the same. The gap between the [Fe/Mg]$_{\rm cc}=-0.5$ and [Fe/Mg]$_{\rm cc}=-0.6$ sequences is also smaller than the gap between [Fe/Mg]$_{\rm cc}=-0.3$ and [Fe/Mg]$_{\rm cc}=-0.4$ or [Fe/Mg]$_{\rm cc}=-0.4$ and [Fe/Mg]$_{\rm cc}=-0.5$. 

For individual stars, the amplitudes $\Acc$ and $\AIa$ can be inferred from the star's [Mg/H] and [Fe/Mg] alone using Equations 13 and 18 of \citetalias{Weinberg2022} \citep[e.g.,][]{Weinberg2019,Hasselquist2024}. In terms of linear ratios, these equations can be written
\begin{equation}\label{eq:acc}
    \Acc = \rm{\{Mg/H\}}
\end{equation}
and 
\begin{equation}\label{eq:aIa}
    \AIa = \Acc\times\left(\frac{\rm{\{Fe/Mg\}-\{Fe/Mg\}_{cc}}}{1-\rm{\{Fe/Mg\}_{cc}}}\right).
\end{equation}
Thus, in this case, the ratio $\AIa/\Acc$ is simply a linear transformation of the linear \{Fe/Mg\} ratio.

Instead of using only [Mg/H] and [Fe/Mg], $\Acc$ and $\AIa$ can also be estimated from a fit to multiple abundances \citepalias[e.g.,][]{Weinberg2022,Griffith2023,Sit2024}. Using multiple elements reduces the effects of random measurement errors in Mg and Fe that may induce artificial correlations in the residual abundances (\citealt{Ting2022}; \citetalias{Weinberg2022}). More specifically, in this work, we adopt the fitting procedure used in \citetalias{Weinberg2022} and \citetalias{Sit2024}---$\chi^2$ minimization to six well-measured APOGEE elements with a range of relative CCSN and SNIa contributions: O, Mg, Si, Ca, Fe, and Ni.

\begin{figure}[!t]
    \centering
    \includegraphics[width=\columnwidth]{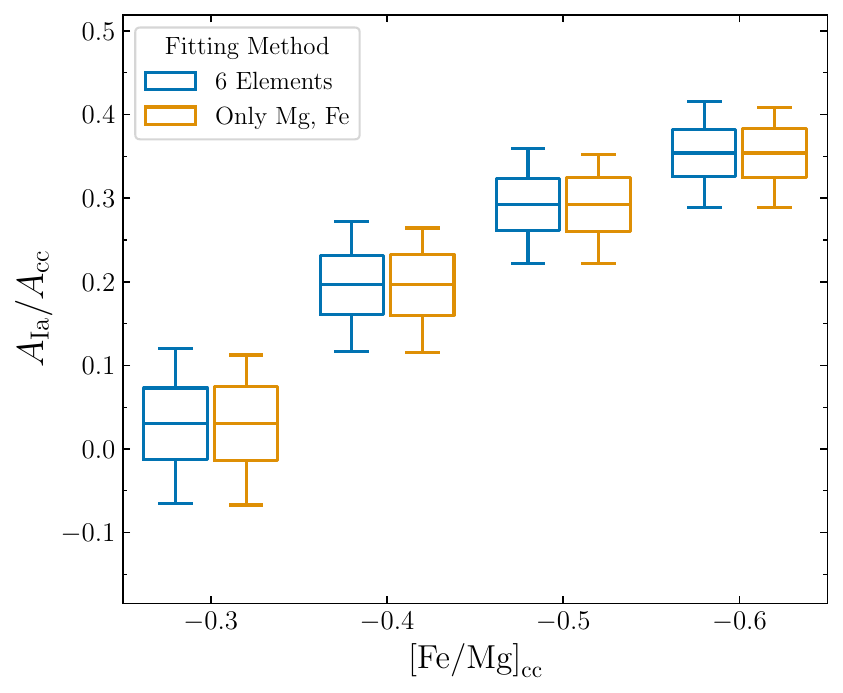}
    \caption{Box-and-whisker plot showing the distribution of the inferred ratio of SNIa to CCSN contribution, $\AIa$/$\Acc$, using the two-process model for different assumed [Mg/Fe]$_{\rm cc}$, for the plateau sample defined in Section \ref{sec:data}. The center horizontal line indicates the median, boxes indicate the interquartile (25th-75th percentile) range, and tails indicate the 5th-95th percentile range of the data. $\Acc$ and $\AIa$ are inferred using Equations \ref{eq:acc} and \ref{eq:aIa} for the orange boxes. $\Acc$ and $\AIa$ are calculated using a $\chi^2$ minimization procedure to the six elements Mg, O, Si, Ca, Fe, and Ni as described in \citetalias{Weinberg2022} and \citetalias{Sit2024} for the blue boxes.}
    \label{fig:Aratio_plateaucompare}
\end{figure}

For stars in the plateau sample (see Section \ref{sec:data}), we use \citetalias{Sit2024} $\log(g)$-calibrated abundances to derive $\Acc$ and $\AIa$. Figure \ref{fig:Aratio_plateaucompare} shows how the distribution of $\AIa/\Acc$ values shifts to higher values as lower $\rm{\{Fe/Mg\}}_{cc}$ values are assumed. The overall $\AIa/\Acc$ distribution at fixed [Fe/Mg]$_{\rm cc}$ is very similar between the Fe and Mg calculation (using Equations \ref{eq:acc}-\ref{eq:aIa}) and the six-element fit; the main notable difference is a slightly longer tail towards high $\AIa/\Acc$ for the six-element fit. The shift from [Fe/Mg]$_{\rm cc}=-0.5$ and [Fe/Mg]$_{\rm cc}=-0.6$ is smaller than [Fe/Mg]$_{\rm cc}=-0.3$ to [Fe/Mg]$_{\rm cc}=-0.4$, similarly to how the $\qIa^X$ and $\qcc^X$ sequences shift in Figure \ref{fig:q_arrays} for Ni, Mn, and Ce. Additionally, there are no variations in $\AIa/\Acc$ correlated with guiding radius $R_{\rm guide}$ or maximum height from the midplane $Z_{\rm max}$ within the plateau sample. These values are taken from the astroNN catalog for DR17\footnote{\url{https://www.sdss4.org/dr17/data_access/value-added-catalogs/?vac_id=the-astronn-catalog-of-abundances,-distances,-and-ages-for-apogee-dr17-stars}}, which uses Gaia eDR3 for distances via \citet[]{LeungBovy2019b} and derives orbital parameters via \citet{Mackereth2018}.

At [Fe/Mg]$_{\rm cc}=-0.4$, the values of $\Acc$ and $\AIa$ are positive over the metallicity and $\alpha$ element range of the plateau sample. This is consistent with \citetalias{Griffith2023} and the physical interpretation that a star cannot have \textit{negative} contribution from a process. Therefore, in the rest of this paper, all two-process quantities calculated for the plateau sample assume [Fe/Mg]$_{\rm cc}=-0.4$. We will show in Section \ref{subsec:linearity} that the choice of [Fe/Mg]$_{\rm cc}$ does not impact the residual abundances (Equation \ref{eq:resid_abund}).

\begin{figure*}[!ht]
    \centering
    \includegraphics[width=\textwidth]{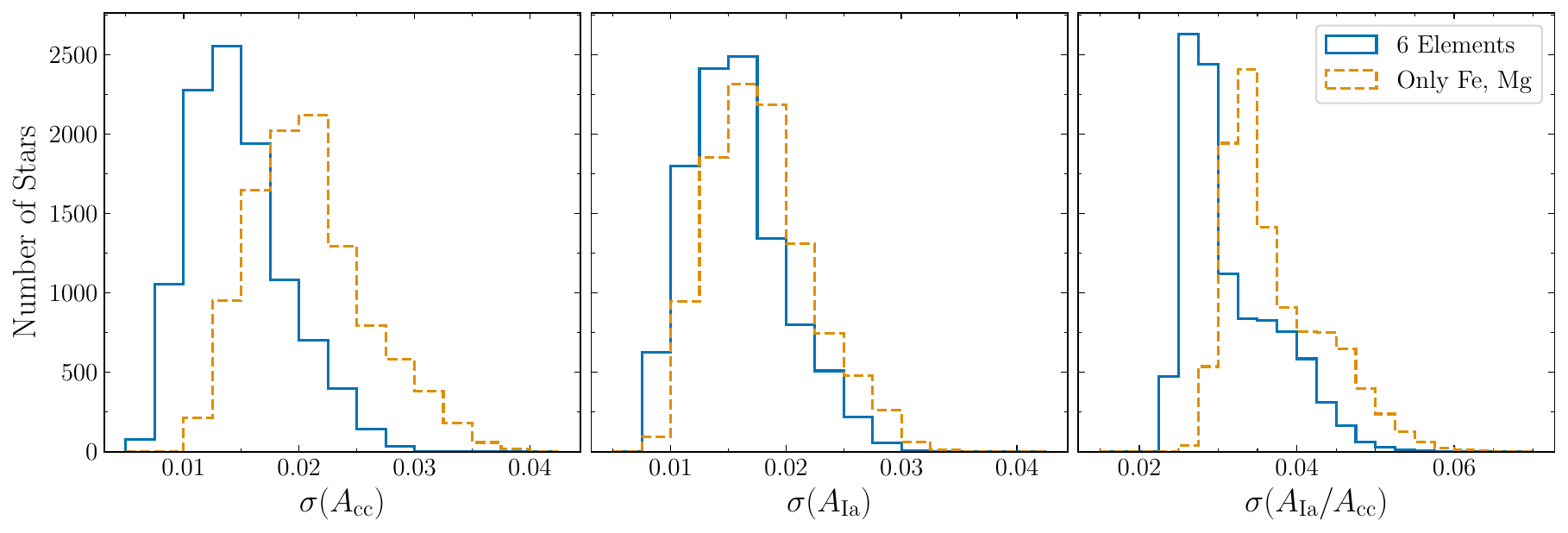}
    \caption{Uncertainty in $\Acc$, $\sigma(\Acc)$ (left); uncertainty in $\AIa$, $\sigma(\AIa)$(middle); and uncertainty in $\AIa/\Acc$, $\sigma(\AIa/\Acc)$ (right) for stars in the plateau sample. For each star, we resample the abundances by drawing from a Gaussian distribution centered at the two-process predicted abundances from the multi-element fit with width corresponding to the observational uncertainty. For each sample, we recalculate $\AIa$ and $\Acc$; blue solid lines indicate that $\AIa$ and $\Acc$ were inferred from the six-element $\chi^2$ minimization procedure, while orange dashed lines indicate that $\AIa$ and $\Acc$ were calculated from only [Mg/H] and [Fe/Mg] using Equations \ref{eq:acc} and \ref{eq:aIa}. The uncertainty $\sigma$ in $\AIa$ and $\Acc$ is the standard deviation of the recalculated $\AIa$ and $\Acc$ over 500 samples. Values were calculated assuming a plateau value of [Fe/Mg]$_{\rm cc}=-0.4$.}
    \label{fig:A_scatter}
\end{figure*}

Figure \ref{fig:A_scatter} shows the distribution of statistical uncertainties in $\AIa$, $\Acc$, and the ratio $\AIa/\Acc$ for the six-element fitting method and the Mg-Fe only method, assuming [Fe/Mg]$_{\rm cc}=-0.4$. For each star, $\sigma(\AIa)$ and $\sigma(\Acc)$ were calculated by resampling the predicted O, Mg, Si, Ca, Fe, and Ni abundances assuming Gaussian reported star-by-star measurement uncertainties, recalculating $\AIa$ and $\Acc$ (using both methods) for each sample, and then calculating the standard deviation of the new $\AIa$ and $\Acc$ values over all samples. The distance between the $\sigma(\AIa)$ and $\sigma(\Acc)$ histogram peaks in Figure \ref{fig:A_scatter} shows that using the six-element fit slightly reduces the uncertainty in the $\AIa$ and $\Acc$ values due to measurement uncertainties, by 0.005-0.01.

\subsection{Residual Abundances and Linearity} \label{subsec:linearity}

\begin{figure*}[!htp]
    \centering
    \includegraphics[width=0.85\textwidth]{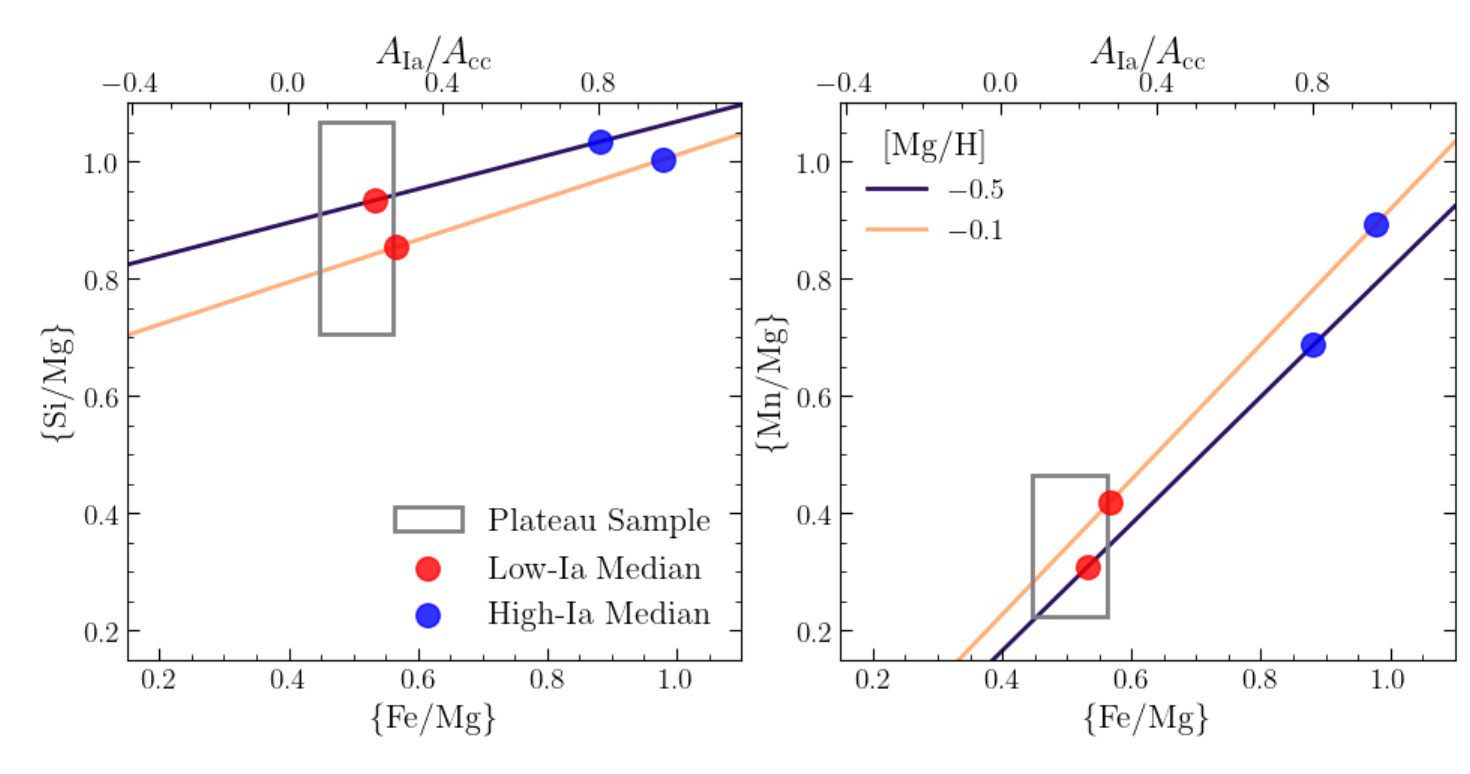}
    \caption{Linear abundances \{Si/Mg\} (left) and \{Mn/Mg\} (right) as a function of \{Fe/Mg\} (bottom axis) or, equivalently, $\AIa/\Acc$ (top axis, using Equation \ref{eq:aIa} assuming [Fe/Mg]$_{\rm cc}=-0.4$). Large red (blue) points indicate the median \{X/Mg\}-\{Fe/Mg\} value of the low-Ia (high-Ia) sequences calculated from the calibration sample at fixed [Mg/H]. The solid lines connect the median points to indicate the two-process predicted linear abundance values. The median points and connecting lines are calculated for $\rm{[Mg/H]} = -0.5$ (dark purple line) and $\rm{[Mg/H]} = -0.1$ (peach line). The gray box encloses the plateau sample: bounds are set by the sample selection in [Fe/Mg] converted to linear abundances and the 0.5 to 99.5 percentile range (99\% of the data) in \{X/Mg\}.}
    \label{fig:linear_interp}
\end{figure*}

From Equations \ref{eq:qcc}-\ref{eq:aIa}, it is straightforward to show that the abundances predicted by the two-process model are
\begin{multline}\label{eq:2proc_pred_linear}
    \rm{\{X/Mg\}_{pred}} = \rm{\{X/Mg\}_{low}}+\left(\rm{\{Fe/Mg\}-\{Fe/Mg\}_{low}}\right)\\
    \times\left(\frac{\rm{\{X/Mg\}_{high}-\{X/Mg\}_{low}}}{\rm{\{Fe/Mg\}_{high}-\{Fe/Mg\}_{low}}}\right).
\end{multline} 
\noindent
When $\qcc^X$ and $\qIa^X$ are inferred from median abundance trends of low-Ia and high-Ia stars, and $\Acc$ and $\AIa$ are inferred from [Mg/H] and [Fe/Mg], the predicted \{X/Mg\} ratios are simply a linear interpolation (or extrapolation) of the median \{X/Mg\} ratios in the low-Ia and high-Ia populations, with \{Fe/Mg\} as the interpolation variable. This mathematically simple result is not obvious from the \citetalias{Weinberg2022} equations because those are expressed in logarithmic abundance ratios [X/Mg] and [Fe/Mg]. Figure \ref{fig:linear_interp} illustrates the linear behavior for two example elements, Si (an $\alpha$ element with primarily CCSN enrichment) and Mn (a Fe-peak element with very significant SNIa enrichment). In this Figure, solid lines connect the high- and low-Ia median points of the calibration sample at fixed [Mg/H] and \textit{are} the predicted linear abundance ratios \{X/Mg\} at that [Mg/H]. The slope of the line represents the extent to which variation in \{X/Mg\} is predicted to vary with \{Fe/Mg\}; it is shallower for a CCSN element like Si but steeper for a SNIa element like Mn in general (though the exact slope and intercept values can change as a function of [Mg/H]).

Most importantly for our purpose, Equation \ref{eq:2proc_pred_linear} shows that the predicted abundances, and thus the residual abundances (Equation \ref{eq:resid_abund}) are \textit{independent} of the assumed value of $\rm{\{Fe/Mg\}}_{cc}$. This result was previously noted by \citetalias{Griffith2023}, who describe it as an ``affine degeneracy" of the two-process model. For different assumed values of \{Fe/Mg\}$_{\rm cc}$, the resulting changes in the process vectors and process amplitudes cancel out exactly so that the predicted abundances remain the same when $\Acc$ and $\AIa$ are calculated from only [Mg/H] and [Fe/Mg].

\begin{figure*}[!t]
    \centering
    \includegraphics[width=\textwidth]{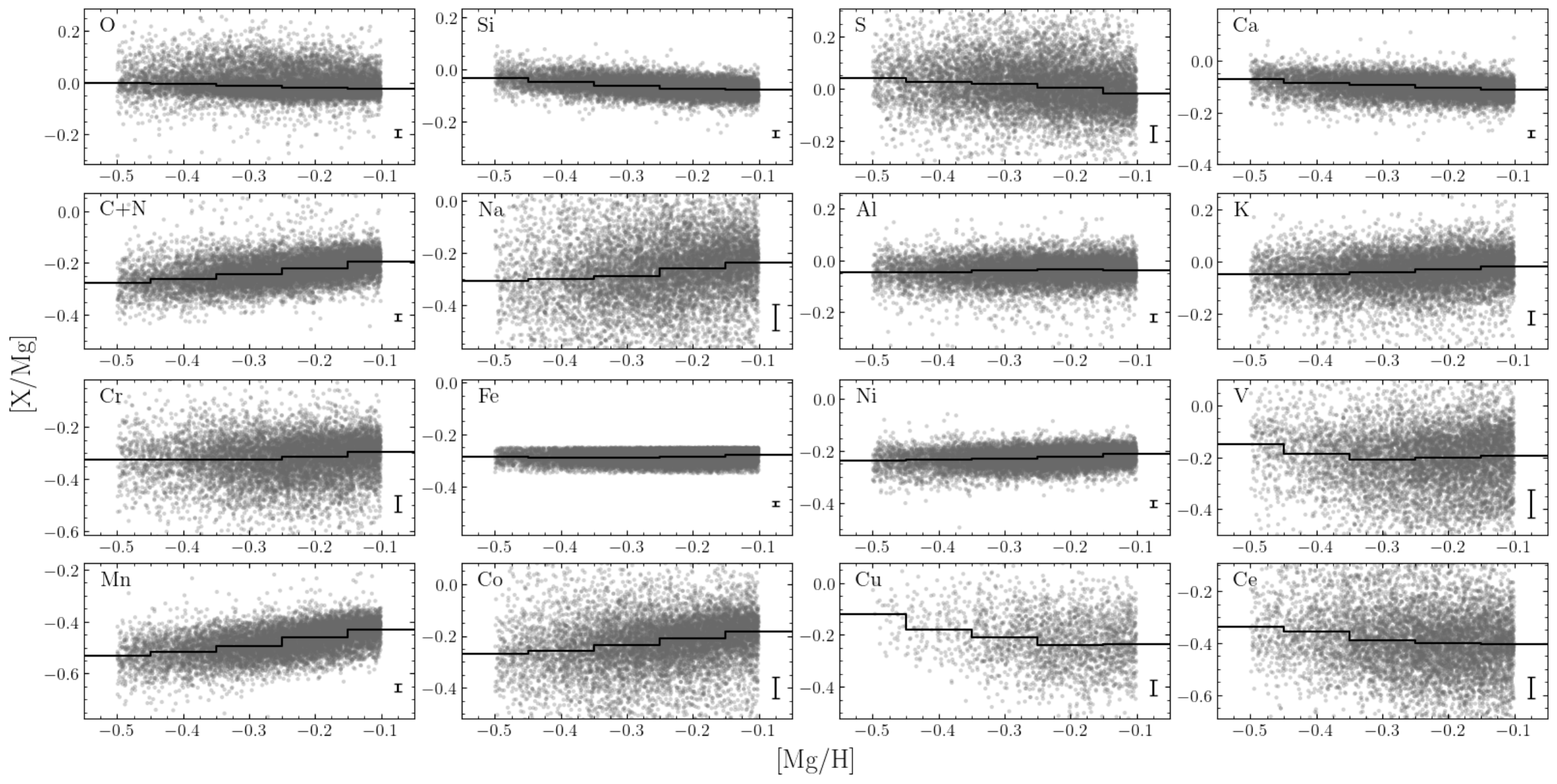}
    \caption{[X/Mg] vs. [Mg/H] for all stars in the plateau sample, for all elements considered in this study (except the reference element Mg). The scale of the y-axis in each panel is the same, shifted only by a zero point to best accommodate the core of each element's distribution, so the level of scatter can be visually compared across all panels. The solid black line indicates the median value of [X/Mg], calculated in 0.1 dex wide [Mg/H] bins. The average observational uncertainty in [X/Mg] for each element, calculated by adding the [X/H] and [Mg/H] uncertainties in quadrature, are shown by the length of the error bars at the bottom right corner in each panel. Observed abundances for Na, Cr, V, Co, and Ce below an upper limit were removed following the procedure outlined in Appendix \ref{appx:upper_limits}.}
    \label{fig:xmg_vs_mgh}
\end{figure*}

We reduce the uncertainty in $\Acc$ and $\AIa$ from measurement uncertainties by using the six-element $\chi^2$-minimization procedure, so the arguments given above are no longer exact. However, the star-by-star changes in $\Acc$, $\AIa$, and residual abundances between a two-element and six-element fit are small. By repeating our full procedure with different [Fe/Mg]$_{\rm cc}$ choices we find that residual abundances are still nearly independent of the adopted value. Therefore, the residual abundances (though not the two-process parameters) presented in the \citetalias{Sit2024} catalog remain valid over all assumptions of [Fe/Mg]$_{\rm cc}$.

\section{Abundance Variation in the Low-Ia Plateau} \label{sec:scatter}

Figure \ref{fig:xmg_vs_mgh} plots [X/Mg] as a function of [Mg/H] for the plateau sample for all elements considered in this work. The overall trend with [Mg/H] is fairly flat for all these elements, but there is significant scatter (except Fe, due to the sample selection criteria). Some of the observed scatter is due to observational uncertainties, whose contribution to the total scatter can be estimated from representative error bars in each panel. Assuming the reported star-by-star measurement uncertainties are correct, the remaining scatter is intrinsic. We wish to understand the origins of this intrinsic scatter, focusing in particular on variation in the SNIa/CCSN ratio.

The elements Na, Cr, V, Co, and Ce exhibited asymmetric tails of the scatter with [X/Mg] for some stars below the median. We found that the majority of stars contributing to this downward scatter reported ASPCAP [X/H] abundance measurements falling below an upper limit given that star's $T_{\rm eff}$ (for Na, V, and Ce) using the upper limit relations presented in \citet{Hayes2022}, or $T_{\rm eff}$ and SNR (for Cr and Co) from the relations of Shetrone et al. (in prep.). The ASPCAP pipeline does not flag upper limits or undetectable abundances. Details on our upper limit flagging procedure and figures showing the affected measurements are presented in Appendix \ref{appx:upper_limits}. The flagged measurements are treated as unreliable and have been removed from all analyses in this Section. Furthermore, we found one extreme outlier star with [Cu/Mg] = 1.06, which is most likely a spurious measurement because examination of the spectrum did not show significant absorption in the known Cu line regions. Therefore, this Cu measurement has also been removed from all subsequent analyses. 

\subsection{Variations Correlated with SNIa/CCSN Ratio}\label{subsec:SNIa_variations}
A plausible source of intrinsic abundance scatter in stars on the low-Ia plateau is remaining variation in the amount of SNIa to CCSN contribution. If this effect is dominant, we would expect that elements with significant SNIa contribution, such as the Fe-peak elements, should vary together, while elements with mostly CCSN contribution, like the $\alpha$ elements, should exhibit little correlated variation. 

We separate the scatter from overall trends by defining the quantity $\Delta$[X/Mg]$_{\rm med}$, the difference between a star's [X/Mg] and the median [X/Mg] of the plateau sample at a similar [Mg/H].The median in [X/Mg] is calculated for the plateau sample in 0.1 dex wide bins centered at [Mg/H] = $-0.5, -0.4, -0.3, -0.2$, and $-0.1$. The medians are shown by the solid black line in each panel in Figure \ref{fig:xmg_vs_mgh}. For a given star, $\Delta$[X/Mg]$_{\rm med}$ is the difference between its measured [X/Mg] and the median [X/Mg] calculated for its [Mg/H] bin, thus removing overall trends with metallicity. $\Delta$[X/Mg]$_{\rm med}$ is defined such that this quantity is positive if the star falls above the median. One can also calculate $\Delta$[X/Mg]$_{\rm med}$ for a predicted abundance, such as from the two-process model. For clarity, we define $\Delta$[X/Mg]$_{\rm med,obs}$ as $\Delta$[X/Mg]$_{\rm med}$ calculated from the observed data, and $\Delta$[X/Mg]$_{\rm med,2proc}$ as $\Delta$[X/Mg]$_{\rm med}$ calculated from two-process predicted abundances.

\begin{figure*}[!th]
    \centering
    \includegraphics[width=0.8\textwidth]{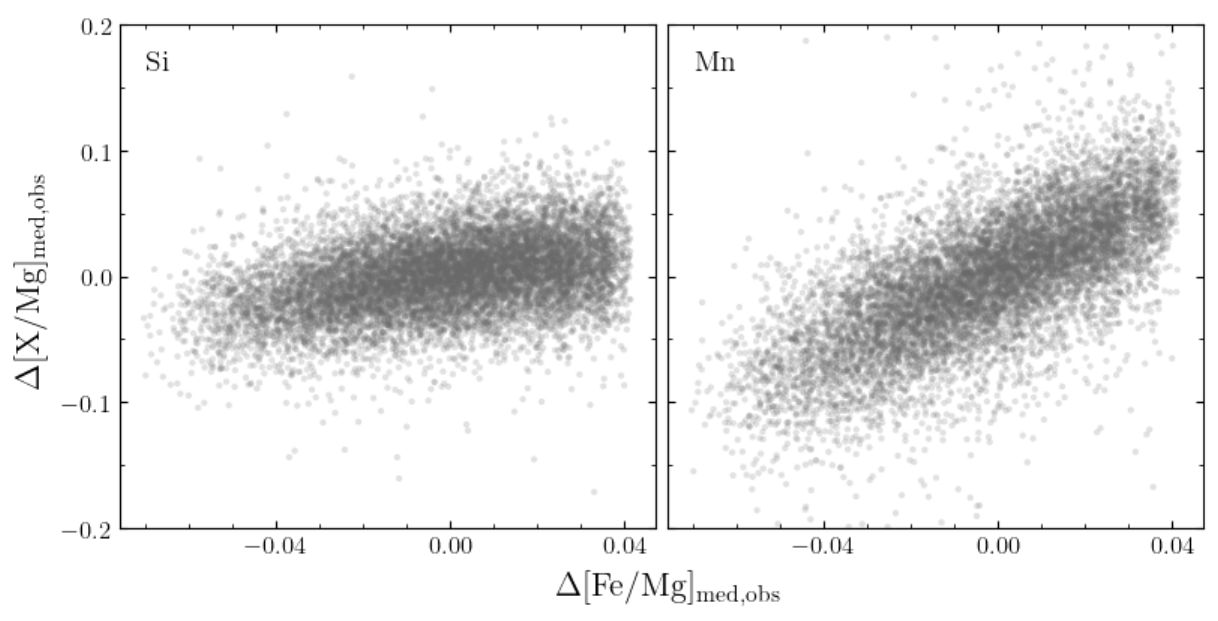}
    \caption{$\Delta$[Si/Mg]$_{\rm med,obs}$ (left) and $\Delta$[Mn/Mg]$_{\rm med,obs}$ (right) as a function of $\Delta$[Fe/Mg]$_{\rm med,obs}$ for the plateau sample. Panels share the same y-axis scale. $\Delta$[Mn/Mg]$_{\rm med,obs}$ shows the stronger correlation with $\Delta$[Fe/Mg]$_{\rm med,obs}$ in this figure because both elements have substantial SNIa contribution, while Si does not.}
    \label{fig:deltaxmg_vs_deltafemg}
\end{figure*}

We show an example of the expected trend in $\Delta$[X/Mg]$_{\rm med}$ between different elements for the full plateau sample in Figure \ref{fig:deltaxmg_vs_deltafemg}. The slope of the $\Delta$[X/Mg]$_{\rm med}$ vs. $\Delta$[Fe/Mg]$_{\rm med}$ is noticeably steeper for Mn than for Si, consistent with with Mn having significantly more SNIa contribution than Si. The slope differences echo that seen in Figure \ref{fig:linear_interp}. We could examine such correlations in the median differences between all the element pairs in our sample, but there is an important caveat in this test: the effect of measurement uncertainty, which adds scatter and can induce spurious correlations because Mg affects both x and y-axis quantities.

In the two-process model fit to only Fe and Mg, the ratio of process amplitudes $\AIa/\Acc$ is a simple linear transformation of linear \{Fe/Mg\} (Equation \ref{eq:aIa}, Figure \ref{fig:linear_interp}). Our six-element fit gives a less noisy estimate of $\AIa/\Acc$, so it is preferable to $\Delta$[Fe/Mg]$_{\rm med,obs}$ when examining the trends with other elements' $\Delta$[X/Mg]$_{\rm med,obs}$.In a small fraction of stars, the six-element fit results in a best-fit $\AIa/\Acc$ that is outside the bounds of the $\AIa/\Acc$ calculated from the linear transformation of {Fe/Mg} (which is set by our sample selection), resulting in a more extended $\AIa/\Acc$ distribution compared to $\Delta$[Fe/Mg]$_{\rm med, obs}$. An additional advantage of incorporating the two-process model is that we can use the predicted abundances to simulate how the elements should correlate if all intrinsic variation is explained by variation in SNIa/CCSN ratio and the only noise is Gaussian with the reported uncertainties.

\begin{figure*}[!pt]
    \centering
    \includegraphics[width=\textwidth]{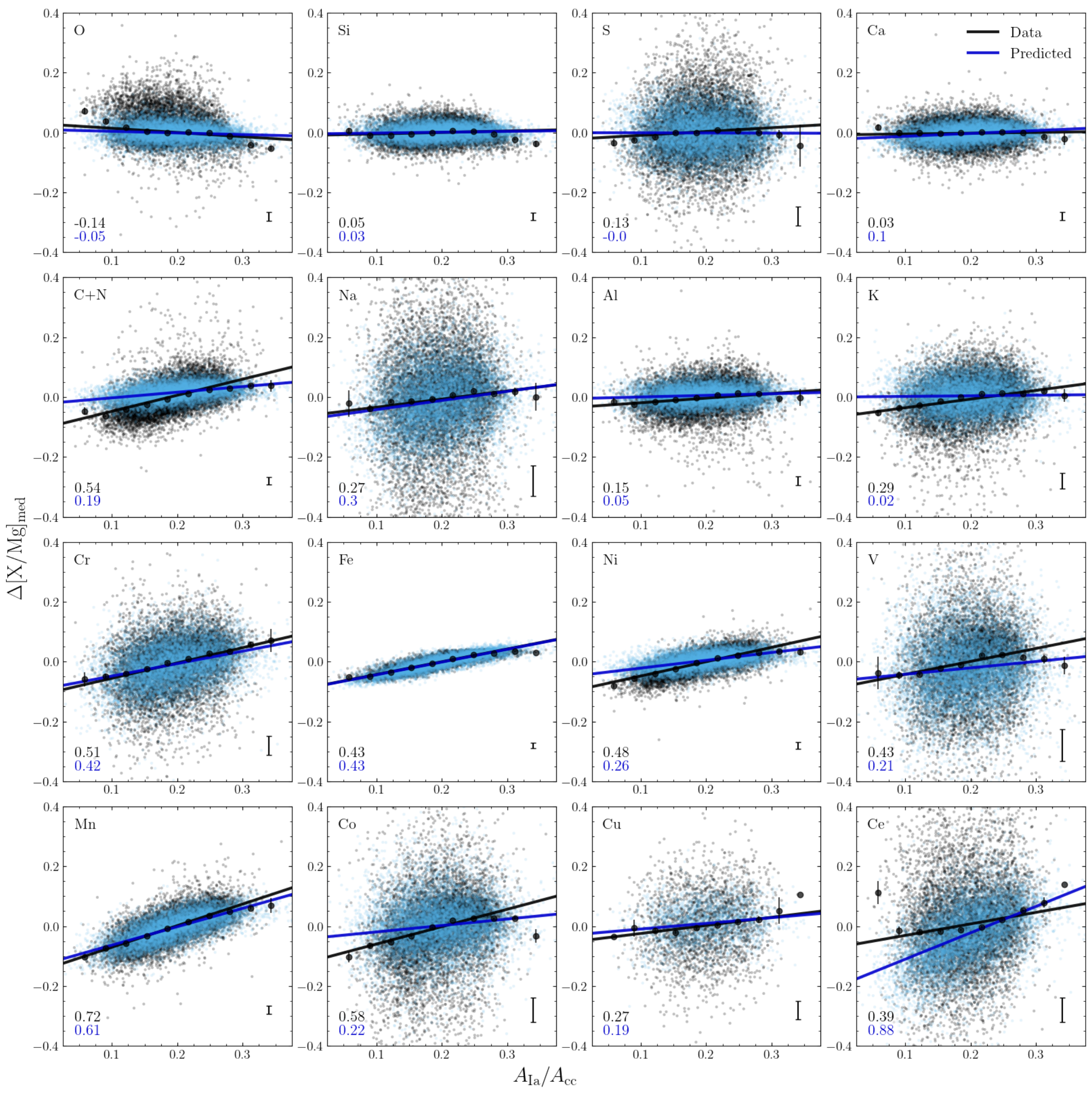}
    \caption{$\Delta$[X/Mg]$_{\rm med}$ vs. $\AIa/\Acc$ for all elements except Mg, our reference element. Small black points are $\Delta$[X/Mg]$_{\rm med,obs}$ values for individual stars (i.e., $\Delta$[X/Mg]$_{\rm med}$ calculated from the observed [X/Mg] of each star), and $\AIa/\Acc$ is calculated from the six-element fit with assumed [Fe/Mg]$_{\rm cc} = -0.4$. Large black points show the median of the black points in 10 evenly sized bins between the minimum and maximum $\AIa$/$\Acc$ values of the plateau sample, and their error bars indicate the uncertainty on each median from 1000 bootstrap resamplings. The thick black line shows the best-fit $\Delta$[X/Mg]$_{\rm med,obs}$ vs. $\AIa/\Acc$ line and the slope value is reported by the black text in the bottom left of each panel. Small blue points show $\Delta$[X/Mg]$_{\rm med,2proc}$ (i.e., $\Delta$[X/Mg]$_{\rm med}$ calculated from the two-process prediction [X/Mg]$_{\rm pred} = \rm{[X/H]}_{\rm pred} - \rm{[Mg/H]}$, where [X/H]$_{\rm pred}$ is given by Equation \ref{eq:2proc_def}, resampled to account for reported observational error) for each star and new $\AIa$ and $\Acc$ values derived from re-fitting to the new simulated [X/H]$_{\rm pred}$. The thick blue line is the best-fit line between $\Delta$[X/Mg]$_{\rm med,2proc}$ and re-fitted $\AIa/\Acc$; the slope value is reported by the blue text in the bottom left of each panel. The length of the error bar on the bottom right of each panel represents the mean observational error for each element. Observed abundances for Na, Cr, V, Co, and Ce below an upper limit were removed following the procedure outlined in Appendix \ref{appx:upper_limits}.}
    \label{fig:linear_fits}
\end{figure*}

Figure \ref{fig:linear_fits} plots $\Delta$[X/Mg]$_{\rm med}$ as a function of the SNIa/CCSN ratio $\AIa/\Acc$. For the observed data ($\Delta$[X/Mg]$_{\rm med,obs}$, gray points in Figure \ref{fig:linear_fits}), $\AIa/\Acc$ is calculated using the two-process model with the six-element fit assuming [Fe/Mg]$_{\rm cc} = -0.4$. We compare this observed multi-element abundance distribution to that predicted by the two-process model, where, by construction, the \textit{only} source of intrinsic scatter is $\AIa/\Acc$. To make this comparison, we generate a sample of predicted abundances by randomly sampling for each star from a Gaussian distribution centered on the two-process predicted [X/Mg] abundance (Equation \ref{eq:2proc_pred_linear}) with standard deviation equal to the reported APOGEE error. We then refit the star's $\AIa$ and $\Acc$ with the perturbed abundances, so that the impact of noise on the two-process parameters is incorporated. Therefore, this sample reflects the predicted distribution of points if the \textit{only} sources of scatter around the median are variations in $\AIa/\Acc$ and measurement error that is Gaussian with the reported uncertainty. The $\Delta$[X/Mg]$_{\rm med,2proc}$ calculated from this [X/Mg] sampling procedure and the corresponding re-fitted $\AIa/\Acc$ are shown in Figure \ref{fig:linear_fits} by the light blue points.

\begin{figure*}[!th]
    \centering
    \includegraphics[width=\textwidth]{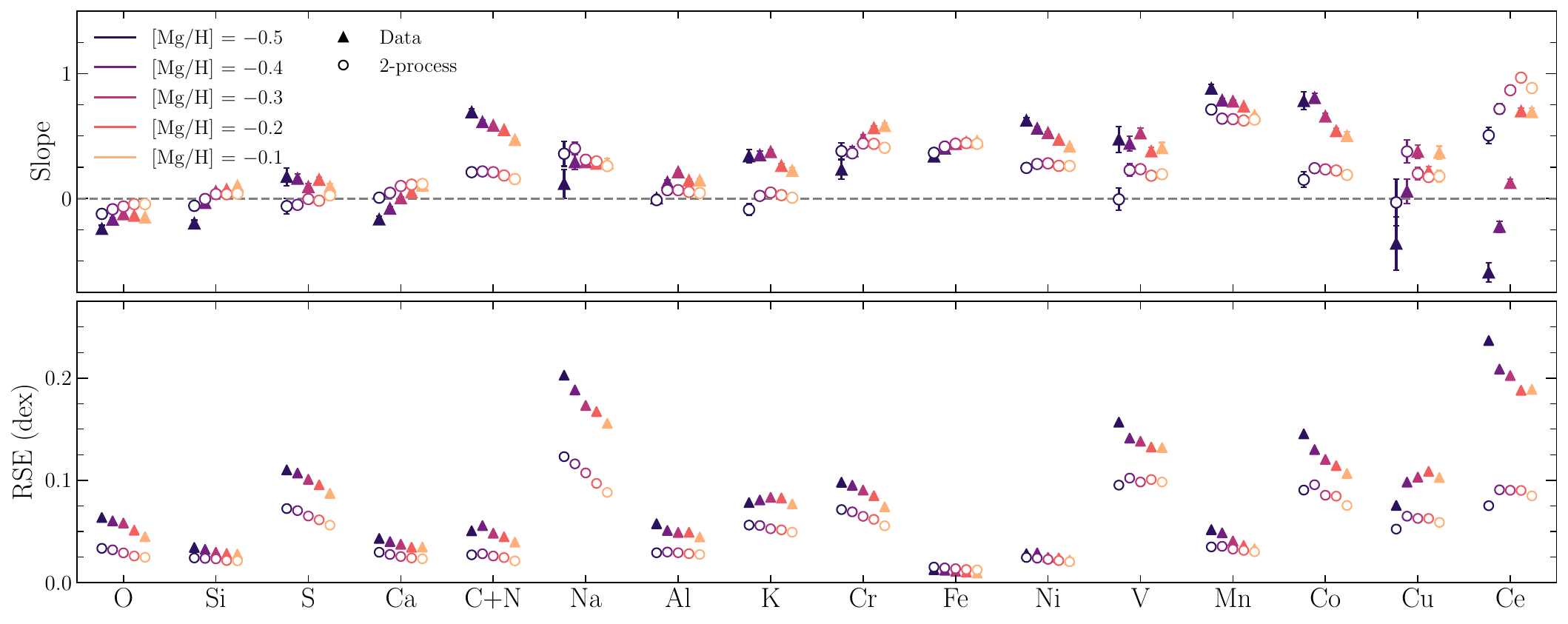}
    \caption{Slopes (top) and residual standard error (RSE, bottom) of the $\Delta$[X/Mg]$_{\rm med}$ vs. $\AIa/\Acc$ plots as illustrated in Figure \ref{fig:linear_fits}. The slope and RSE are calculated in 0.1 dex [Mg/H] bins centered at [Mg/H] = $-0.1, -0.2, -0.3, -0.4, -0.5$, to show the potential metallicity dependence of these quantities. Error bars in the top panel indicate the uncertainty on the slope from the linear regression covariance matrix; they are frequently smaller than the marker. The RSE in the bottom panel is calculated as the standard deviation of residuals from the linear fit. In both panels, filled triangles indicate the data (Figure \ref{fig:linear_fits}, black lines) and the open circles indicate the two-process prediction with Gaussian observational errors (Figure \ref{fig:linear_fits}, blue lines). Observed abundances for Na, Cr, V, Co, and Ce below an upper limit were removed prior to calculation of the slope and RSE following the procedure outlined in Appendix \ref{appx:upper_limits}.}
    \label{fig:slopes+RSE}
\end{figure*}

The slope of the $\Delta$[X/Mg]$_{\rm med}$ vs. $\AIa/\Acc$ relation quantifies the SNIa contribution to scatter on the low-Ia plateau. Slopes are calculated from a linear regression using inverse-variance weighting of the observational errors. The best-fit line calculated over all metallicities is shown in Figure \ref{fig:linear_fits} by the solid lines corresponding to the colored points used for the regression. In Figure \ref{fig:slopes+RSE}, we instead perform the linear regression in individual 0.1 dex [Mg/H] bins (the same as those used to calculate the medians and $\Delta$[X/Mg]$_{\rm med}$) rather than over the entire plateau sample. The top panel of Figure \ref{fig:slopes+RSE} shows these slopes for all elements as a function of metallicity. For most elements, the slopes do not show notable metallicity dependence, so the trend seen for the full plateau sample in Figure \ref{fig:linear_fits} is representative. We note that for Si, the slope seen in Figure \ref{fig:deltaxmg_vs_deltafemg} appears significantly shallower in the observed slope in Figure \ref{fig:linear_fits}, and Figure \ref{fig:slopes+RSE} has a $\Delta$[Si/Mg]$_{\rm med,obs}$ vs. $\AIa/\Acc$ slope of $\approx$0. This apparent slope change is partly due to the larger axis scale in Figure \ref{fig:linear_fits}, but it arises primarily because some of the deviations are incorporated into $\AIa/\Acc$ from the fitting procedure. We would expect similar effects in the other fitting elements (O, Ca, Fe, and Ni).

In general, the Fe-peak elements like Cr, Fe, Ni, V, and Mn, which are theoretically expected to have significant contribution from SNIa, exhibit steeper $\Delta$[X/Mg]$_{\rm med,obs}$ vs. $\AIa/\Acc$ slopes. Conversely, elements with primarily CCSN contribution, like the $\alpha$ elements O, Si, and Ca and the odd-Z elements Al and K, have observed slopes close to 0. The positive slopes in only elements with known significant SNIa contribution imply that variation in the amount of SNIa enrichment \textit{is} a significant source of scatter on the low-Ia plateau. Therefore, the median trend through the low-Ia plateau does \textit{not} represent ``pure" CCSN enrichment, though the CCSN yield ratios could lie at the upper end of the observed scatter.

The slopes are similar between the observed data and two-process predictions for most elements, indicating that the two-process model predicts SNIa contribution on the plateau well. The two elements with the largest differences between their observed and predicted slopes are Ce and C+N. The slope calculated with $\Delta$[Ce/Mg]$_{\rm med,2proc}$ is steeper than the slope calculated with $\Delta$[Ce/Mg]$_{\rm med,obs}$, indicating that the two-process model over-predicts the SNIa contribution to Ce. Because the delayed contribution to Ce probably reflects AGB enrichment rather than SNIa enrichment, it is not surprising that the linear trend inferred from the low-Ia and high-Ia medians does not apply within the low-Ia population (see Figure \ref{fig:linear_interp}). Within our plateau sample, [Ce/Mg] exhibits little trend with $\AIa/\Acc$, but large scatter.

Conversely, the $\Delta$[(C+N)/Mg]$_{\rm med,2proc}$ slope is shallower than the $\Delta$[(C+N)/Mg]$_{\rm med,obs}$ slope. As with Ce, we expect the delayed contribution to C and N to come from AGB enrichment rather than SNIa enrichment, so it is not surprising that the linear trend between the low-Ia and high-Ia medians does not describe the behavior within the low-Ia population. However, it is notable that the prediction is too steep for Ce but too shallow for C+N. This different behavior could possibly arise from different metallicity dependence of C, N, and Ce yields, but we have no full explanation for these trends. We suspect, but do not know, that most of the variation in [(C+N)/Mg] within our plateau sample comes from variation in birth [N/Mg], because CCSN are expected to be the primary sources of C while AGB stars make a large and metallicity-dependent contribution to N enrichment \citep[see, e.g.,][]{Andrews2017,Rybizki2017,Johnson2023}. The departures of C+N and Ce trends from those predicted based on the low-Ia and high-Ia median may provide clues to the physics of AGB enrichment in the early MW disk. 

Other elements with small slope discrepancies include K, V, and Co, but these elements have large observational uncertainties, and non-Gaussian abundance errors or incorrect error estimates might affect the slopes. Finally, Ni also shows a small discrepancy between observed and predicted slopes. This is unexpected because Ni is considered a well-measured element in ASPCAP that we used in the six-element two-process fit and is not known to have any nucleosynthetic contributions from processes outside of SNIa and CCSN. When NLTE effects are accounted for, the ratio of Ni to Fe increases as metallicity decreases at subsolar metallicities \citep{Eitner2023}; this could possibly be a related signature. If the discrepancy is real, a possible explanation of the steeper Ni slope is that the population-averaged Ni yields are higher at plateau metallicities. A potential cause of different yields may be differences in the SNIa progenitor landscape. Ni, Mn, and Cr yields are known to be different between sub-Chandrasekhar and near-Chandrasekhar mass SNIa \citep{Kobayashi2020b}. Therefore, if the ratio of sub-Chandrasekhar to near-Chandrasekhar white dwarfs is different at the time of plateau formation compared to the current MW disk, this would result in different SNIa yields on the plateau even at same metallicity. However, as the Mn (and to a lesser extent, Cr) yields are also affected in this scenario, we would expect to also see a discrepant slope in Mn, but the difference between the $\Delta$[Mn/Mg]$_{\rm med,obs}$ and $\Delta$[Mn/Mg]$_{\rm med,2proc}$ slopes is smaller than for Ni.

\subsection{Scatter Around the SNIa/CCSN Correlation} \label{subsec:scatter}

While the slope quantifies the extent of $\AIa/\Acc$ variation on the plateau, the scatter around the linear fits is also interesting. We observe in Figure \ref{fig:linear_fits} that several elements have a more extended distribution in the data (gray points) than is predicted by the two-process model plus observational errors (blue points). This suggests that there may be an additional source of intrinsic variation in that element that is unrelated to the SNIa/CCSN ratio, or that the reported measurement errors are underestimated.

We measure the scatter around the best-fit lines using the residual standard error (RSE, the standard deviation of residuals from the best-fit line). The bottom panel of Figure \ref{fig:slopes+RSE} plots the RSE as a function of metallicity for all elements. Since the RSE is calculated from the best-fit line, it takes into account potential differences in the $\Delta$[X/Mg]$_{\rm med}$ vs. $\AIa/\Acc$ slopes between the observed data and two-process predictions. The measurement uncertainty is not explicitly included in the RSE calculation, though inverse-variance weighting is used for the linear regression to fit the line. Because the two-process prediction points incorporate observational scatter, a larger RSE for the data implies variation not explained by changes in the $\AIa/\Acc$ ratio plus Gaussian measurement uncertainties.

Broadly, most of the more well-measured elements (e.g., Si, Ca, Al, Fe, Ni, Mn) show similar RSE between the data and prediction, indicating that measurement noise is sufficient to explain the remaining variation after contributions from a varying SNIa/CCSN ratio are removed. In other words, the amount of SNIa contribution and observational errors are the primary sources of variation on the plateau for these elements. Many of the elements that have larger observational uncertainties (e.g., S, Na, K, Cr, V, Co, Cu, and Ce) have higher observed RSE than predicted RSE. The explanation may be that the reported measurement uncertainties are still underestimated, or that the errors have non-Gaussian tails, since several of these elements (most notably S, Na, V, Cu, and Ce) have relatively weak lines that are often affected by blending. In these cases, the shallower spectral features make abundances difficult to measure and less reliable, particularly at lower abundances, which may increase the scatter. In some cases, especially Ce and perhaps C+N and Na, the larger observed RSE could reflect intrinsic scatter in an enrichment process that is neither CCSN nor SNIa.

O is a fairly well-measured element with small observational uncertainties, but it has a slightly elevated observed RSE compared to its prediction. Looking at the corresponding panel of Figure \ref{fig:linear_fits}, this may be explained by the upward splash of gray points towards lower $\AIa/\Acc$, which is not well captured by the more symmetric scatter of the blue points. Although the reason is unclear, the $\Delta$[O/Mg]$_{\rm med,obs}$ distribution is noticeably asymmetric: $\approx$7\% of the stars have $\Delta$[O/Mg]$_{\rm med,obs}>0.1$ while $\approx$1\% of stars have $\Delta$[O/Mg]$_{\rm med,obs}<-0.1$ (recall that $\Delta$[X/Mg]$_{\rm med}$ is defined such that the distribution is centered at 0). Similarly, C+N has a small population of stars with higher $\Delta$[(C+N)/Mg]$_{\rm med,obs}$ in Figure \ref{fig:linear_fits}. We use the reported ASPCAP [C/Fe] measurement error as the total C+N uncertainty because C generally dominates over N by number (see \citetalias{Weinberg2022} and \citetalias{Sit2024}), so C+N may also be affected by underestimated measurement uncertainty. Alternatively, the high $\Delta$[(C+N)/Mg]$_{\rm med,obs}$ points could be a physical population affected by AGB mass transfer. Notably, while Ce has much larger observational errors than C+N, there is also a small population of stars with elevated $\Delta$[Ce/Mg]$_{\rm med,obs}$. We have have investigated these outlier populations and found that many of these stars are elevated in both $\Delta$[(C+N)/Mg]$_{\rm med,obs}$ and $\Delta$[Ce/Mg]$_{\rm med,obs}$, a likely signature of mass transfer, as found for the broader disk population by \citetalias{Weinberg2022} (see their Figure 14). 

For those elements where there is a metallicity trend at all, the RSE decreases with increasing [Mg/H]. For the predicted RSE, this is consistent with metal lines becoming stronger and easier to measure at higher metallicity and thus the observational uncertainties decreasing. The most notable exception to this trend is Cu, which shows generally higher RSE at higher [Mg/H], but only $\approx$20\% of the stars in our sample have a Cu measurement at all, and there are fewer measurements in the low [Mg/H] bins.

The size of the difference, if any, between the observed and predicted RSE is also broadly the same across [Mg/H] bins. Ce is a significant exception to this, as the difference between its observed and predicted RSE is largest at low [Mg/H]. The observational uncertainty could simply be more severely underestimated at lower metallicities. However, as Ce is a primarily AGB element, there could be a more physical variation in the amount of AGB enrichment relative to SNIa enrichment at lower metallicities. In such a scenario, different regions in the Galaxy may have different levels of AGB contribution at a fixed metallicity due to varying star formation histories and the time-dependence of the AGB yields. As metal-poor stars are born from gas that experienced fewer enrichment events overall, they are more impacted by fluctuations in the level of AGB enrichment and the extent of gas mixing (or lack thereof) in their birth region. 

\begin{figure*}
    \centering
    \includegraphics[width=\textwidth]{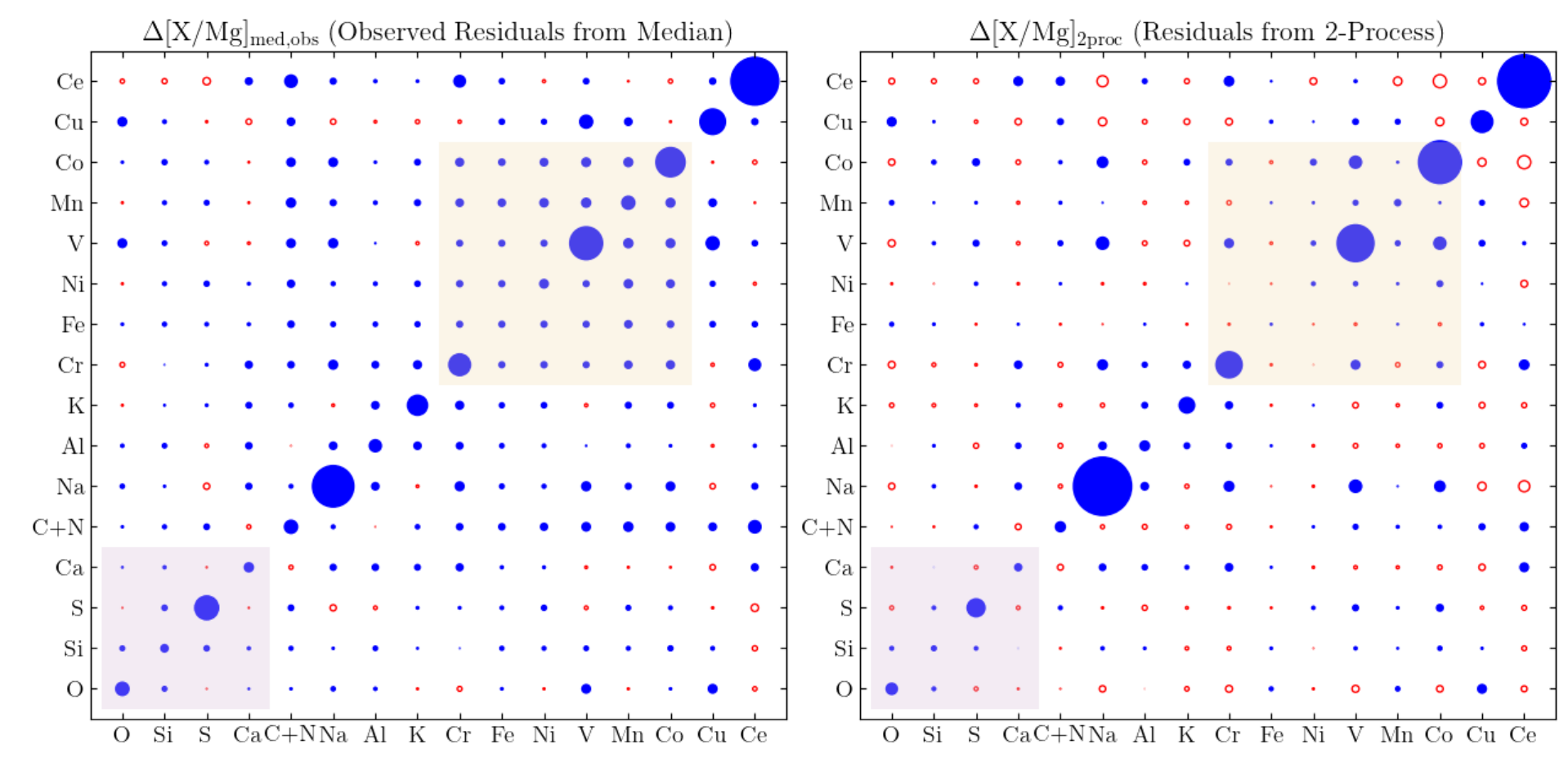}
    \caption{Covariance matrix of residuals $\Delta$[X/Mg] for stars on the low-Ia plateau. In the left panel, $\Delta$[X/Mg] are the observed residuals from the median plateau value, i.e., $\Delta$[X/Mg]$_{\rm med,obs}$ as in Figure \ref{fig:linear_fits}. In the right panel, $\Delta$[X/Mg] are the two-process residual abundances as described in Equation \ref{eq:resid_abund}, but with Mg used as the reference element (Equations \ref{eq:2proc_deltaxmg1}-\ref{eq:2proc_deltaxmg2}), and the covariance from measurement uncertainty has been subtracted. The $\alpha$ (Fe-peak) elements are highlighted by the purple (yellow) square in both panels. Blue filled circles and red open circles denote positive and negative values, respectively, with area proportional to the value of the covariance. Circles in both panels are on the same scale; for reference, on the left panel, the Na-Na circle is 0.03, O-O is 0.003, and Na-K is 0.000003.}
    \label{fig:covar_dot_plot}
\end{figure*}

\subsection{Covariances Between Elements}
Correlations can be measured robustly even when observational uncertainties are comparable in magnitude to the intrinsic dispersion \citep{Ting2022}. Furthermore, the residual correlation structure can encode information about additional enrichment pathways (beyond SNIa and CCSN, which are accounted for by the two-process model), stochastic sampling of the interstellar medium (ISM), and potentially, interesting stellar evolution pathways such as binary mass transfer. We compute the covariance for each element pair $ij$ as
\begin{multline}\label{eq:covar}
    c_{ij} = \langle\ (\Delta[{\rm X}_i/{\rm Mg}]-\langle\Delta[{\rm X}_i/{\rm Mg}]\rangle)\ \times\\
    (\Delta[{{\rm X}_j/{\rm Mg}]-\langle\Delta[{\rm X}_j/{\rm Mg}]\rangle)}\ \rangle.
\end{multline}
As we are calculating the covariances of precomputed \textit{residuals}, the expectation value in Equation \ref{eq:covar}, $\langle\Delta[{\rm X/Mg}]\rangle$, is the mean of the residuals over the entire plateau sample which should be close to, but may not be exactly, 0. As the covariances are calculated for element pairs, the exclusion of any one element does not affect the presence of correlations (or lack thereof) between the remaining elements; similarly, the inclusion of a new element only adds the potential for correlations with that new element without affecting the others.

Figure \ref{fig:covar_dot_plot} shows the covariance matrices for the observed residuals from the median, $\Delta$[X/Mg]$_{\rm med,obs}$, in the left panel and the covariance of residuals from the two-process model, $\Delta$[X/Mg]$_{\rm 2proc}$, in the right panel. The size of each dot corresponds to the magnitude of the covariance, and the diagonals represent the variance of the residuals in each element. Blue filled dots represent positive covariances while open red dots represent negative covariances.

The two-process model returns predictions in [X/H] (Equation \ref{eq:2proc_def}) such that the residual abundances (Equation \ref{eq:resid_abund}) are defined as $\Delta$[X/H]$_{\rm 2proc}$. To use Mg as the reference element, we use the following equation for the $\Delta$[X/Mg]$_{\rm 2proc}$ residuals:
\begin{align}
    \Delta\rm{[X/Mg]_{\rm 2proc}} ={}& 
    (\rm{[X/H]}_{\rm meas}-\rm{[Mg/H]}_{\rm meas})\nonumber\\
    &-(\rm{[X/H]}_{\rm 2proc}-\rm{[Mg/H]}_{\rm 2proc})\label{eq:2proc_deltaxmg1}\\
    ={}&\Delta\rm{[X/H]_{\rm2proc}}-\Delta\rm{[Mg/H]_{\rm 2proc}}.\label{eq:2proc_deltaxmg2}
\end{align}
\noindent $\Delta$[X/H]$_{\rm 2proc}$ is defined in Equation \ref{eq:resid_abund} and $\Delta$[Mg/H]$_{\rm 2proc}$ is Equation \ref{eq:resid_abund} with X = Mg. As in \citetalias{Weinberg2022}, for the two-process residual covariance matrix (right panel), we subtract the covariance from measurement uncertainty, so that variation from \textit{only} the two-process model remains. We also computed the covariance matrix of \textit{predicted} median deviations, i.e. $\Delta$[X/Mg]$_{\rm med,2proc}$ (blue points in Figure \ref{fig:linear_fits}), but do not show it in Figure \ref{fig:covar_dot_plot} as all values are positive (i.e., the two-process model predicts that all residuals from the median should vary together to some extent).

The left panel of Figure \ref{fig:covar_dot_plot} has many positive covariances, indicating that most elements tend to deviate from the median in the same direction. In particular, the Fe-peak elements, highlighted by the yellow square, have significant positive residuals; this is consistent with our previous result that variation in SNIa/CCSN contribution is a significant source of correlated abundance scatter on the low-Ia plateau. The broadly positive residuals within an element group are also seen, though to a lesser extent, in the $\alpha$ elements highlighted by the purple box. We excluded the ``Fe-cliff" element Cu from the Fe-peak yellow square here because it behaves differently from the other Fe-peak elements: Cu has (small) negative covariances with Cr and Co. Interestingly, \citet{Kobayashi2020} find that Co and Cu are both largely produced by hypernovae, so one might expect these two elements in particular to be correlated.

The two-process model captures the variation in the SNIa/CCSN ratio, so that the covariance in the two-process residuals should have the covariance from SNIa/CCSN variation removed. The right panel of Figure \ref{fig:covar_dot_plot} shows the covariance matrix of the two-process residuals. As expected, the signs of the two-process residual covariances are much more mixed, compared to the mostly-positive median residual covariances. The magnitude of the covariances between Fe-peak elements have significantly decreased in the two-process residuals, and some have even changed sign, compared to the median residuals. However, V, Mn, and Co, which are odd-Z Fe-peak elements, still show some positive correlation with each other in the two-process residuals. The $\alpha$ elements already have small covariances, and their magnitudes are not significantly different between the median and two-process residuals. However, the Ca-O and Ca-Si covariances have different signs. Similarly, the light odd-Z elements like C+N, Na, and Al do not show any obvious correlations with each other in their two-process residuals, though some covariance signs change from the median residuals. The two-process residuals of C+N are less correlated with those of the Fe-peak elements than the median residuals and are still positively correlated with Ce. Since C+N has contribution from a delayed process, this result indicates that this delayed process is not fully explained by the same delayed process (SNIa) producing the Fe-peak elements; furthermore, as Ce is an AGB element, this suggests that an AGB process also contributes to C+N.

\citetalias{Weinberg2022} removed outliers with a two-process residual greater than 10 times the measurement error for that element from their covariance calculations. If we also perform this cut, we find similar results to before, but the most affected element is Ce. More specifically, the Ce-(C+N) intrinsic $\Delta$[X/Mg]$_{\rm 2proc}$ covariance and Ce-Al $\Delta$[X/Mg]$_{\rm med,obs}$ covariance change sign from positive to negative, while the Ce-Ni, Ce-Mn, and Ce-Co $\Delta$[X/Mg]$_{\rm med,obs}$ covariances change from negative to positive. The rms scatter of Ce (diagonal element) also decreases. The majority of the removed outliers appear to be the same population of stars with strongly enhanced Ce and C+N found in \citetalias{Weinberg2022}, who propose that they are rare products of AGB mass transfer, where extra C+N and Ce from the AGB star are deposited onto the observed star's photosphere. Their presence here indicates that such mass transfer events occur in stars on the low-Ia plateau.

\section{Discussion} \label{sec:conclusions}

In this work, we investigate the origin of multi-element abundance scatter of stars in the low-Ia (high-$\alpha$) plateau of the MW disk, with a particular focus on the effects of variation in the ratio of SNIa to CCSN enrichment. Our test is based on the idea that elements with large SNIa contribution should vary together while the variation in elements with little SNIa contribution should be smaller and less correlated. One could simply examine the correlations between deviations from the median trends across many element pairs, but we use a two-process model to more precisely estimate the SNIa/CCSN ratios and to predict what the plateau variations should look like if, indeed, the only sources of scatter are differences in SNIa/CCSN ratio and measurement uncertainty.

First, we tested the effect of the assumed Fe-to-Mg ratio from pure CCSN enrichment, [Fe/Mg]$_{\rm cc}$, on the two-process model as formalized by \citetalias{Weinberg2022}. When the process amplitudes $\AIa$ and $\Acc$ are inferred from only [Mg/H] and [Fe/Mg], the predicted \textit{linear} \{X/Mg\} abundance is simply a linear interpolation (or extrapolation) of the median \{X/Mg\} ratios in the high- and low-Ia populations along \{Fe/Mg\}. It directly follows that the predicted abundances are independent of the assumed [Fe/Mg]$_{\rm cc}$, and we show that this independence holds to an excellent approximation even when using a multi-element $\chi^2$ fit for $\AIa$ and $\Acc$.

Because of the linearity of the two-process model, the assumed pure-CCSN plateau value [Fe/Mg]$_{\rm cc}$ only scales the process vectors and amplitudes, and it does so in a way that results in the same predicted abundances and without affecting the quality of the fit. This implies that the \textit{absolute} scale of the $\qIa$ and $\qcc$ process vectors, which can set empirical constraints on the IMF-averaged CCSN and SNIa yields \citep[e.g.,][]{Griffith2021b}, is poorly constrained. On the other hand, because the predicted abundances themselves do not change, we also demonstrate that residual abundance analyses---namely, the residual abundance trends in different populations (e.g., \citetalias{Weinberg2022,Sit2024}; \citealp{Hasselquist2024}) and with Galactic position/kinematics \citep{Griffith2024}---are robust to the [Fe/Mg]$_{\rm cc}$ model assumption.

Next, we use the derived process amplitudes and predicted abundances from the two-process model to investigate elemental abundance scatter on the plateau. We examine $\Delta$[X/Mg]$_{\rm med}$, the [X/Mg] residuals from median trends in the plateau, as a function of $\AIa/\Acc$, the ratio of SNIa/CCSN enrichment inferred from the two-process model. Our main result is that these values are positively correlated, especially in the Fe-peak elements, indicating that variation in the amount of SNIa enrichment is a significant contributor to the [X/Mg] scatter on the low-Ia plateau. This interpretation is further supported by the Fe-peak elements having lower covariances in the two-process residuals than the median residuals.

Our results clearly establish that there is a systematic difference in the SNIa/CCSN enrichment ratio between stars at the ``top" of the low-Ia plateau and stars at the ``bottom," a range of roughly 0.1 dex in [Mg/Fe]. This rules out the idea that the median [Mg/Fe] on the low-Ia plateau represents the core collapse yield ratio [Mg/Fe]$_{\rm cc}$, with abundance ratios scattering stochastically above and below because of, e.g., stochastic sampling of the CCSN population. It remains possible that [Mg/Fe]$_{\rm cc}$ lies at the upper boundary of the plateau, at [Mg/Fe]$_{\rm cc} = 0.35-0.4$ (See Figure \ref{fig:sample}). However, our results also lend credence to the idea that the true CCSN ratio is substantially higher, perhaps [Mg/Fe]$_{\rm cc}\approx0.6$, and the observed low-Ia plateau already includes substantial SNIa enrichment \citep{Maoz2017,Conroy2022}. Maintaining a nearly flat plateau at this intermediate [Mg/Fe] requires accelerating star formation if the ISM is well mixed \citep{Conroy2022,Chen2023}, though a similar effect may be achieved if early enrichment is spread throughout the circumgalactic medium and only slowly returns to the star-forming ISM \citep{Mason2024}.

For most elements we find good agreement between the predicted and observed slope of $\Delta$[X/Mg]$_{\rm med}$ vs. $\AIa/\Acc$, but there are exceptions. Most notably, the observed slope is steeper than predicted for C+N and shallower than predicted for Ce. For both abundances, we expect the delayed contribution comes from AGB stars rather than SNIa, so disagreement with the two-process model is not too surprising, but we do not know why the sign of the discrepancy is different in these two cases. The intermediate mass stars that dominate AGB yields have lifetimes that are typically shorter than the $\sim$1 Gyr median delay of the SNIa enrichment, which can vary from 0.18 to 2.5 Gyr depending on the model (see Table 2 of \citealt{Dubay2024}).\footnote{\citet{Dubay2024} find that delay time distributions with fewer prompt SNIa, which have have median delay times of $\sim$1$-$2.5 Gyr, result in multi-zone chemical evolution models that better fit the [$\alpha$/Fe] distribution in the MW disk.} The steeper trend of C+N is arguably the more natural result because of the larger change in C+N over the restricted time range of the thick disk compared to the interval between the thick and thin disks. Ni also has a slightly steeper observed than predicted trend, which suggests that the Ni/Fe yield ratio of SNIa may be slightly higher in the low-Ia population.

The scatter about the $\Delta$[X/Mg]$_{\rm med}$ vs. $\AIa/\Acc$ trend is higher for most elements than we predict based on the two-process model plus Gaussian scatter with the reported measurement error. In some cases, this excess scatter may indicate that the measurement uncertainties are underestimated, or at least that there are non-Gaussian tails to the error distribution. Recent studies have been mixed on how accurate the ASPCAP uncertainties are for DR17: \citet{Mead2024} found that the uncertainties are underestimated at dwarf galaxy metallicities, while \citet{Sinha2024} found that the uncertainties were sufficiently well-estimated in open clusters near solar metallicity. The plateau sample analyzed in this work falls in between these two metallicity ranges. Future work on calibrating ASPCAP abundance uncertainties with high-resolution spectra will shed more light on this issue. Accurate abundance measurement uncertainties are essential for meaningful work on investigating the intrinsic abundance scatter of any population, not just the low-Ia plateau. Nonetheless, our main result that differences in the SNIa/CCSN ratio are a significant cause of variation on the low-Ia plateau holds regardless of potential changes to the magnitude or distribution of measurement errors because it is based on the slope of the $\Delta$[X/Mg]$_{\rm med}$ vs. $\AIa/\Acc$ trend, not the scatter around it.

Some of the excess scatter may be intrinsic. In particular, if AGB (or other delayed enrichment channels) make an important contribution to an element, then variation in the AGB/SNIa ratio at a given SNIa/CCSN ratio will lead to variation in that abundance (see \citealt{Griffith2022} for Y and Ba in GALAH). Covariances of residual abundances can help to identify channels that affect multiple elements, but they become difficult to interpret when the scatter is only moderately larger than the measurement errors, which is typically the case for residual abundances in APOGEE.

Much of the attention in MW chemical evolution modeling has gone to understanding the difference in abundance trends between the thin and thick disks, and the bimodality of the [$\alpha$/Fe] distribution at a given [Fe/H]. These features remain incompletely understood, with a variety of possible explanations. Extending earlier studies of the [$\alpha$/Fe] scatter within the low-Ia population \citep{deLis2016,Vincenzo2021}, our results show that there are significant variations of the SNIa/CCSN ratio \textit{within} the low-Ia population at fixed [Mg/H], producing coherent variations in the abundance ratios of many elements. However, the source of the variation itself is not obvious; potential causes may include stellar migration and/or uneven mixing of the ISM during the formation of the plateau. They further show distinctive behavior of some elements expected to have delayed contribution from AGB stars. Asteroseismic ages of APOGEE stars in the Kepler field show that the thick disk stars have a remarkably small age range, at least at the solar annulus \citep{Pinsonneault2024}. Producing the observed abundance variations within a short span of time is a challenge for models of the early evolution of the MW disk.

\section{Acknowledgments}
We thank Chris Hayes for providing upper limit relations and helpful discussion, Sten Hasselquist for detailed comments that helped improve the paper, and Yuan-Sen Ting for interesting discussion on the two-process model. We also thank the anonymous referee for helpful feedback.

TS acknowledges support from the NSF Graduate Research Fellowship Program under Grant No. DGE-2240614.

E.J.G. is supported by an NSF Astronomy and Astrophysics Postdoctoral Fellowship under award AST-2202135.

This work is supported by NSF grant AST-2307621.

Funding for the Sloan Digital Sky 
Survey IV has been provided by the 
Alfred P. Sloan Foundation, the U.S. 
Department of Energy Office of 
Science, and the Participating 
Institutions. 

SDSS-IV acknowledges support and 
resources from the Center for High 
Performance Computing  at the 
University of Utah. The SDSS 
website is www.sdss4.org.

SDSS-IV is managed by the 
Astrophysical Research Consortium 
for the Participating Institutions 
of the SDSS Collaboration including 
the Brazilian Participation Group, 
the Carnegie Institution for Science, 
Carnegie Mellon University, Center for 
Astrophysics | Harvard \& 
Smithsonian, the Chilean Participation 
Group, the French Participation Group, 
Instituto de Astrof\'isica de 
Canarias, The Johns Hopkins 
University, Kavli Institute for the 
Physics and Mathematics of the 
Universe (IPMU) / University of 
Tokyo, the Korean Participation Group, 
Lawrence Berkeley National Laboratory, 
Leibniz Institut f\"ur Astrophysik 
Potsdam (AIP),  Max-Planck-Institut 
f\"ur Astronomie (MPIA Heidelberg), 
Max-Planck-Institut f\"ur 
Astrophysik (MPA Garching), 
Max-Planck-Institut f\"ur 
Extraterrestrische Physik (MPE), 
National Astronomical Observatories of 
China, New Mexico State University, 
New York University, University of 
Notre Dame, Observat\'ario 
Nacional / MCTI, The Ohio State 
University, Pennsylvania State 
University, Shanghai 
Astronomical Observatory, United 
Kingdom Participation Group, 
Universidad Nacional Aut\'onoma 
de M\'exico, University of Arizona, 
University of Colorado Boulder, 
University of Oxford, University of 
Portsmouth, University of Utah, 
University of Virginia, University 
of Washington, University of 
Wisconsin, Vanderbilt University, 
and Yale University.

\software{
Matplotlib \citep{matplotlib}, astropy \citep{astropy2013,astropy2018,astropy2022}, NumPy \citep{numpy}, SciPy \citep{scipy}, pandas \citep{pandas1,pandas2}
}

\appendix
\section{Upper Limits}\label{appx:upper_limits}

\begin{figure}[!th]
    \centering
    \includegraphics[width=\linewidth]{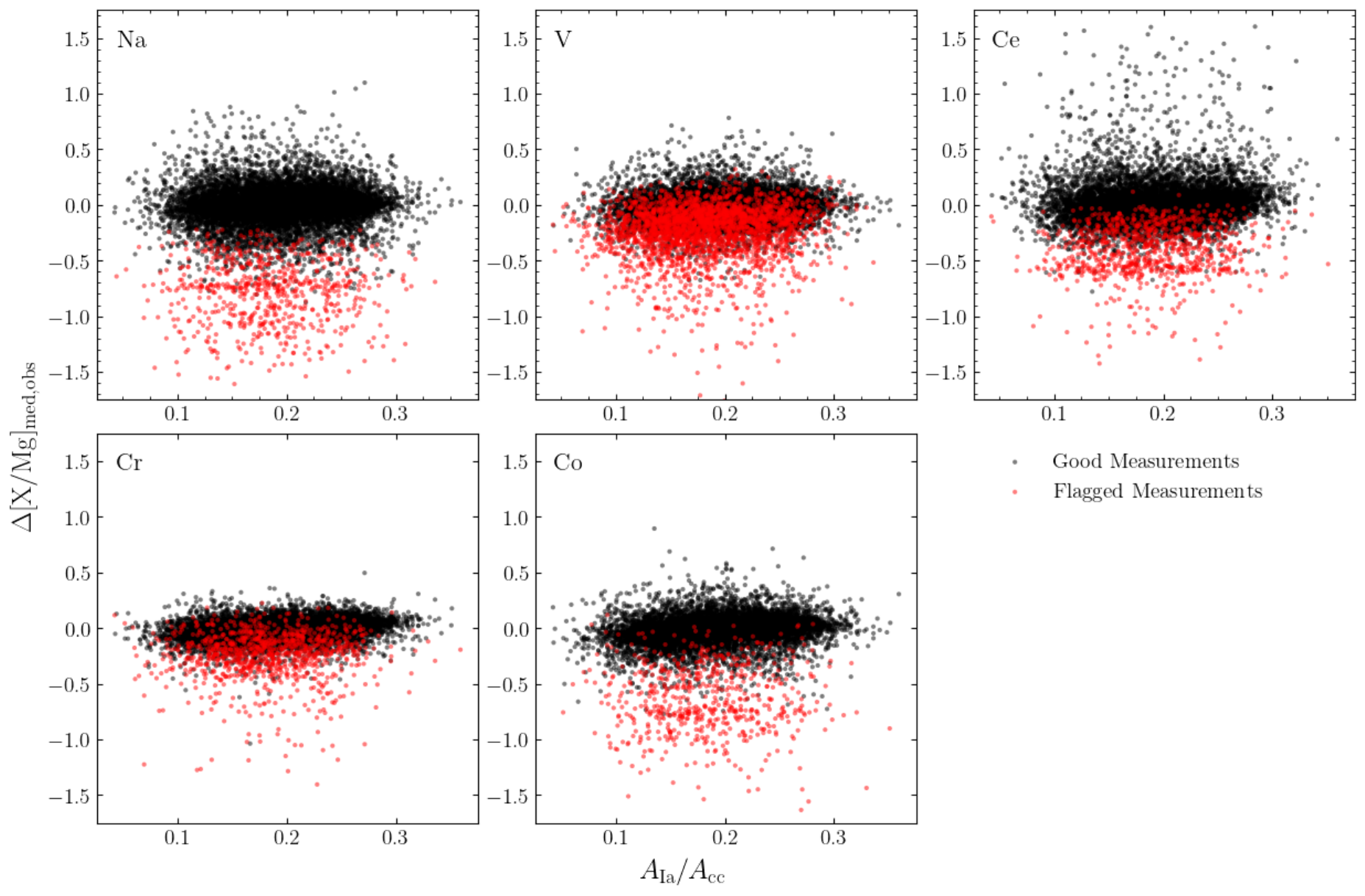}
    \caption{$\Delta$[X/Mg]$_{\rm med,obs}$ vs. $\AIa/\Acc$, as in Figure \ref{fig:linear_fits}, for elements significantly affected by upper limit removal (Na, V, Ce, Cr, and Co). Elements using the BAWLAS upper limit relations \citep{Hayes2022} are in the top row while elements using the relations derived for dSph (Shetrone et al., in prep.) are in the bottom row. Measurements falling below the calculated upper limit for a given star's $\Teff$ (Na, V, Ce) or $\Teff$ and SNR (Cr, Co) and subsequently removed from analyis in Section \ref{sec:scatter} are highlighted in red. The remaining stars are plotted in black.}
    \label{fig:upper_limit_removal}
\end{figure}

As ASPCAP does not naturally return upper limit estimates, we must turn to external analyses to determine the minimum possible abundance that can be measured in a given star. If the abundance measurement for a star falls below that limit, we assume that the spectral line was not strong enough for a detection and ASPCAP was likely fitting noise, and thus remove that measurement from analyses in Section \ref{sec:scatter}. Figure \ref{fig:upper_limit_removal} highlights the stars removed due to this upper limit analysis in red.

Because our sample is selected for high SNR like the BAWLAS catalog \citep{Hayes2022}, we prioritize the BAWLAS upper limit relations for the available overlapping elements. The BAWLAS upper limit relations are derived by synthesizing synthetic spectra over a grid of $\Teff$ and element abundance, measuring where the line depths are some percent of the continuum value, and fitting a linear function of $\Teff$ at fixed continuum threshold. We select the relations corresponding to a 1\% continuum threshold. Then, for each star and each element, we calculate the upper limit of each line at that star's $\Teff$, then take the lowest line-by-line limit as the element's overall upper limit (so the line used can change with $\Teff$). The elements most affected by this analysis are Na, V, and Ce, which all also showed an asymmetric scatter to [X/Mg] below the median. We find that $\approx$7\% of the total Na measurements, $\approx$26\% of the V measurements, and $\approx$10\% of the total Ce measurements fall below upper limits; these fractions increase to $\gtrsim$75\% for Na and $\gtrsim$90\% for V and Ce of all measurements when we consider only stars where $\Delta$[X/Mg]$_{\rm med,obs}<-0.4$ dex. While $\approx$18\% of the total O measurements also fall below upper limits, the flagged measurements have approximately the same distribution as the unflagged measurements and their removal does not noticeably impact any analyses presented in Section \ref{sec:scatter}, so we do not remove them.

For elements not included in the BAWLAS catalog, we use the upper limit relations from Shetrone et al., in prep. Like BAWLAS, these relations were derived using a grid of synthetic spectra at a range of abundances and $\Teff$. For each element, a weighted average of the line depth was measured for all individual lines; then, a function linear in $\Teff$ and quadratic in $\log_{10}({\rm SNR})$ was fit to the detection thresholds, defined where the average line depth is 4 times the noise per pixel. We note an important caveat that these relations were derived for analysis of APOGEE dwarf spheroidal galaxies with lower metallicities and SNR than analyzed here; however, no other upper limits for APOGEE's spectra for non-BAWLAS elements are currently available. Using this set of upper limits, we find a significant impact on Cr and Co: $\approx$10\% of the Cr measurements and $\approx$6\% of the Co measurements fell below the detection threshold. Like Na, V, and Ce, Cr and Co also showed asymmetric scatter to low [X/Mg] prior to upper limit removal, and $\gtrsim$90\% of Cr measurements and $\gtrsim$80\% of Co measurements with $\Delta$[X/Mg]$_{\rm med,obs}<-0.4$ dex were below upper limits. Additionally, there were 7 flagged K measurements (0.06\%), but we do not remove them from the Section \ref{sec:scatter} analysis.

Overall, removal of these upper limits results in a decrease in observed RSE of $\approx$0.1 dex for Na, $\approx$0.05 dex for V, $\approx$0.03 dex for Ce, $\approx$0.04 dex for Cr, and $\approx$0.09 dex for Co. For Na and Co, this RSE decrease exceeds the median observational error. The nature of the flagged measurements mostly having $\Delta$[X/Mg]$_{\rm med,obs}\lesssim-0.4$ also generally results in the $\Delta$[X/Mg]$_{\rm med,obs}$--$\AIa/\Acc$ slopes decreasing (becoming flatter, as these five affected elements all have positive slopes). 

\section{Reproducibility}

We have produced a public Git repository containing a Jupyter notebook that reproduces all figures in this paper. Included with the repository are also data tables for $q$ values assuming different [Fe/Mg]$_{\rm pl}$ (for reproducing Figure \ref{fig:q_arrays}) and upper limits of the plateau sample in all elements (for reproducing Figure \ref{fig:upper_limit_removal}). The repository is accessible at \url{https://github.com/tawnysit/plateau_variation}.

\bibliography{plateau_scatter_bib}{}
\bibliographystyle{aasjournal}

\end{document}